\documentclass[prd,onecolumn,notitlepage,eqsecnum,nofootinbib,floatfix]{revtex4-1}
\usepackage{amssymb}
\usepackage{amsmath}
\usepackage{amsfonts} 
\usepackage{bm}
\usepackage{graphicx}
\newcommand{\gothf}{\mathfrak{f}}
\allowdisplaybreaks
\begin{document}
\title{Self-force from conical singularity, without renormalization}   
\author{Michael LaHaye and Eric Poisson}  
\affiliation{Department of Physics, University of Guelph, Guelph,
  Ontario, N1G 2W1, Canada} 
\date{April 28, 2020} 
\begin{abstract} 
We develop an approach to calculate the self-force on a charged particle held in place in a curved spacetime, in which the particle is attached to a massless string and the force is measured by the string's tension. The calculation is based on the Weyl class of static and axially symmetric spacetimes, and the presence of the string is manifested by a conical singularity; the tension is proportional to the angular deficit. A remarkable and appealing aspect of this approach is that the calculation of the self-force requires no renormalization of the particle's  electric field. This is in contract with traditional methods, which incorporate a careful and elaborate subtraction of the singular part of the field. We implement the approach in a number of different situations. First, we examine the case of an electric charge in Schwarzschild spacetime, and recover the classic Smith-Will force in addition to a purely gravitational contribution to the self-force. Second, we turn to the case of electric and magnetic dipoles in Schwarzschild spacetime, and correct expressions for the self-force previously obtained in the literature. Third, we replace the electric charge by a scalar charge, and recover Wiseman's no-force result, which we generalize to a scalar dipole. And fourth, we calculate the force exerted on extended bodies such as Schwarzschild black holes and Janis-Newman-Winicour objects, which describe scalarized naked singularities.   
\end{abstract}  
\maketitle

\section{Introduction and summary} 
\label{sec:intro} 

In a classic work, Smith and Will \cite{smith-will:80} calculated the self-force acting on an electric charge held in place in the Schwarzschild spacetime of a nonrotating black hole. In flat spacetime, the electric field lines emanating from the charge would be isotropically distributed around the particle, and the net force on the charge would vanish (this in spite of the infinite value of the field at the particle's position). In a curved spacetime, however, the electric field is modified by the spacetime curvature, the field lines are no longer isotropic, and the net force no longer vanishes. Smith and Will relied on an expression for the electric field provided by Copson \cite{copson:28} and corrected by Linet \cite{linet:76}. Because the field diverges at the charge's position, an essential aspect of their calculation was a careful regularization of this singular field, followed by a renormalization to a finite piece that is solely responsible for the self-force. 

The Smith-Will self-force is given by $F_{\rm self} = q^2 M/r_0^3$, where $q$ is the particle's electric charge, $M$ the mass of the black hole, and $r_0$ the charge's radial position (in the usual Schwarzschild coordinates). The self-force points away from the black hole, and therefore represents a repulsive effect. Their result was generalized to electric charges in the Reissner-Nordstr\"om spacetime \cite{zelnikov-frolov:82}, to scalar charges \cite{wiseman:00, burko:00b, burko-liu:01}, and to higher-dimensional black holes \cite{frolov-zelnikov:12a, frolov-zelnikov:12b, beach-poisson-nickel:14, taylor-flanagan:15, harte-flanagan-taylor:16}. The Smith-Will self-force is nearly universal, in the sense that its expression is largely independent of the internal composition of the gravitating body \cite{drivas-gralla:11, isoyama-poisson:12}. 

The calculation of the self-force by Smith and Will leaves a number of questions unanswered. Among these are: What is the external agent responsible for holding the charge at its fixed position? Isn't this agent a significant source of gravitation? What is the impact of the electric field on the spacetime geometry? Can these modifications to the gravitational field alter the description of the self-force? Should there not also be a gravitational component to the self-force? And what is the precise operational meaning of the self-force? 

Our aim with this paper is to provide a more complete description of the self-force and to supply answers to these questions. We have two main concerns. The first is the nature of the external agent: what is holding the charge at its fixed position? The second is the operational meaning of the self-force: who is measuring this force, and where is it being measured? As we start providing answers to these questions, we shall find that the other queries find answers as well. 

We first make a choice of external agent. We declare that the charge shall be prevented from falling into the black hole by being attached to a massless string. The string extends from the particle to infinity, where it is held firmly by an observer (to be thought of as an actual person). The force required of this observer is equal to the tension in the string, and this force accounts for the local acceleration of the charge in the Schwarzschild spacetime, along with all self-force effects. This choice of external agent provides at once an operational meaning for the self-force: it is measured by the string's tension, after subtracting off the local acceleration. 

\begin{figure} 
\includegraphics[width=0.9\linewidth]{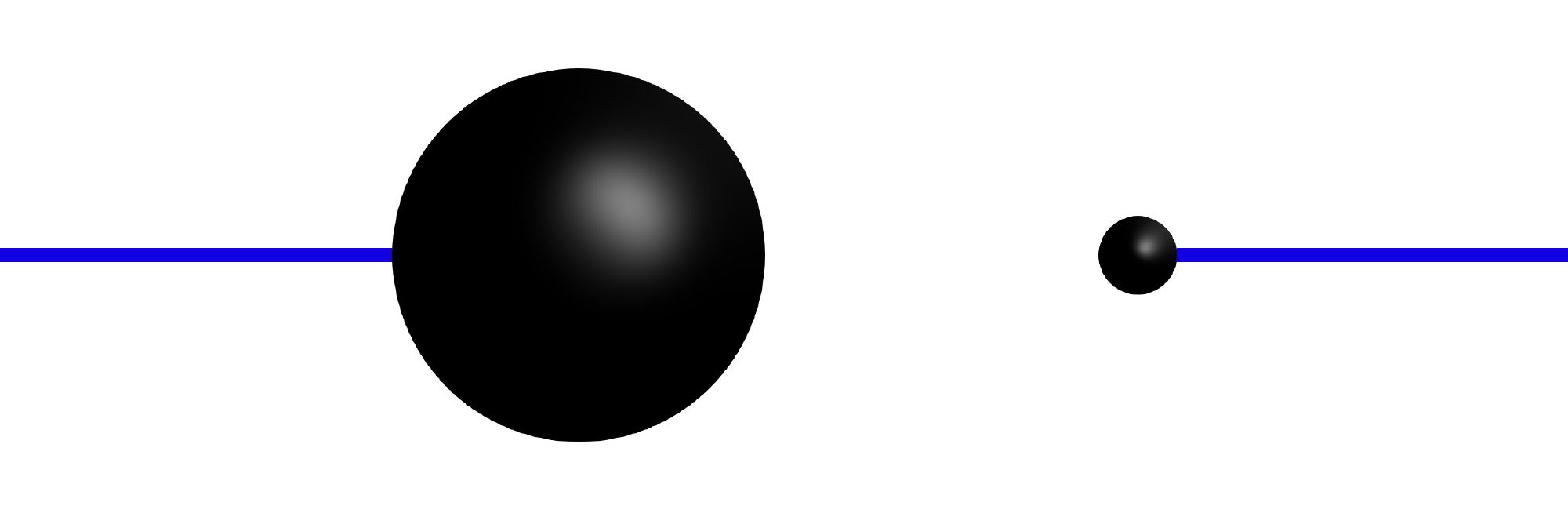}
\caption{A massless string is attached to a particle (right) to prevent it from falling toward a black hole. A second string is attached to the black hole (left) to prevent it from falling toward the particle.}   
\label{fig:fig1} 
\end{figure}

The black hole must also be prevented from falling toward the particle. We do this by attaching a second massless string to the black hole. This string extends from the black hole to infinity, where it is held by a second observer. Newton's third law guarantees that the force required of this observer, which is equal to the tension in the second string, is equal to the force supplied by the first observer. The situation is depicted in Fig.~\ref{fig:fig1}. 

To make all this precise, we work with the class of static and axially symmetric spacetimes described by the Weyl metric (see, for example, Ch.~10 of Ref.~\cite{griffiths-podolsky:09}) 
\begin{equation} 
ds^2 = -e^{-2U}\, dt^2 + e^{2U} \bigl[ e^{2\gamma}(d\rho^2 + dz^2) + \rho^2\, d\phi^2 \bigr],  
\label{weyl-intro} 
\end{equation} 
in which $U$ and $\gamma$ are gravitational potentials that depend on $\rho$ and $z$. The class includes the Schwarzschild spacetime, whose metric can easily be recast in this form, and it includes also all the spacetimes to be considered in this paper, which are modifications to the Schwarzschild spacetime that account for the field created by the particle. This class of metrics is very convenient to work with, because $U$ satisfies a Poisson-type equation for which a multitude of solutions have been found, and because $\gamma$ can then be obtained from $U$ by evaluating quadratures. 

In the metric of Eq.~(\ref{weyl-intro}), the ratio of proper circumference to proper radius for a small circle $\rho = \mbox{constant}$ around the $z$-axis is given by $2\pi \exp(-\gamma^{\rm axis})$, where $\gamma^{\rm axis} := \gamma(\rho=0,z)$ is the value of $\gamma$ on the axis. Elementary flatness demands that this ratio be precisely $2\pi$, and for this we must have $\gamma^{\rm axis} = 0$. Failure to achieve this implies that an angular deficit measured by $\beta := 2\pi[1 - \exp(-\gamma^{\rm axis})]$ has been introduced in the geometry; the spacetime contains a conical singularity. This singularity signals the presence of a material source on the axis, which can be interpreted as a Nambu-Goto string, a one-dimensional object whose mass density $\mu$ is equal to its tension $T$ --- the string traces a two-dimensional world sheet in spacetime. The string is massless because its sole gravitational manifestation is the nonzero $\gamma^{\rm axis}$; it makes no contribution to $U$. The string's tension is given by \cite{vilenkin:81, hiscock:85, gott:85, linet:85} 
\begin{equation} 
T = \frac{\beta}{8\pi} = \frac{1}{4} \Bigl[ 1 - \exp\bigl(-\gamma^{\rm axis} \bigr) \Bigr], 
\label{tension-intro} 
\end{equation} 
and an observer holding this string at infinity would need to exert a force $F = T$. 

These observations define our strategy in this paper. We consider a charged particle held in place by a massless string in the spacetime of a nonrotating black hole. We calculate the electric field produced by this charge, and we calculate the gravitational potential $U$ for this system, going well beyond the test-charge approximation in which the metric is kept to its Schwarzschild expression. Next we find $\gamma^{\rm axis}$, calculate the string's tension according to Eq.~(\ref{tension-intro}), and thus obtain the force required of an observer at infinity to hold the string. In doing all this, we manage to answer all the questions listed previously.  

In addition to providing a precise operational meaning to the force and answers to these questions, a substantial advantage of the method developed in this paper is that {\it the calculation of the force requires no regularization and no renormalization of the particle's electric field}, which is badly singular at the particle's position. Because the force follows from the angular deficit instead of an evaluation of the field acting on the charge, there is absolutely no need to deal with the singular nature of the field. In our opinion, the absence of renormalization in this scheme is a most powerful conceptual advance over the traditional methods of calculation.\footnote{We hasten to point out that in spite of this bold claim, our calculations are not entirely free of regularization. While it is true that there is no need to regularize the electric field (a big deal for us), we shall nevertheless have to contend with other infinities. The first occurs when we attempt to define the particle's mass at second order in perturbation theory, because at this order the mass incorporates the particle's gravitational binding energy, which is formally infinite for a point mass. The second occurs when we attempt to introduce a redshift factor for photons emitted at the particle and received at infinity; this is infinite because the gravitational potential of a point mass is infinite at the particle's position. In both cases we require a mild form of regularization to remove the infinities. For the particle's mass, we absorb the binding energy within the definition of the mass at second order. For the redshift, we subtract the particle's contribution from the total effect. All this is to be contrasted with the complete absence of regularization for the electric field.}     

We are not the first authors to exploit the Weyl metric of Eq.~(\ref{weyl-intro}) to calculate the force required to keep gravitating objects at fixed positions. There is, in fact, a vast literature on this topic, reviewed in Ch.~10 of Ref.~\cite{griffiths-podolsky:09}, which features strings and struts to keep the objects still. An interesting subset of this literature \cite{barker-oconnell:77,  kimura-ohta:77, bonnor:81, tomimatsu:84, bonnor:93, azuma-koikawa:94, perry-cooperstock:97, breton-manko-sanchez:98, bini-geralico-ruffini:07, alekseev-belinkski:07} is concerned with the equilibrium of two (or more) massive and charged objects. This line of inquiry culminated in the discovery of a class of solutions to the Einstein-Maxwell equations \cite{manko:07, manko-ruiz-sanchezmondragon:09} that describes two charged black holes held apart by a strut. In recent papers \cite{krtous-zelnikov:19a, krtous-zelnikov:19b}, Krtou\v{s} and Zelnikov investigated the thermodynamic and self-force aspects of these spacetimes.    

The literature reviewed in the preceding paragraph is concerned mostly with {\it exact solutions} to the field equations, and it considers situations in which $\gamma$ {\it can be obtained globally}. To investigate the conical singularity and calculate the string's tension, however, it is not necessary to know $\gamma$ everywhere in the spacetime. It suffices to know $\gamma^{\rm axis}$, the value of $\gamma$ on the $z$-axis. Giving up on a global $\gamma$ and focusing instead on $\gamma^{\rm axis}$ should open up the way for much more exploration.  

In this paper we develop techniques that allow us to calculate $\gamma^{\rm axis}$ directly, by exploiting the singular nature of all fields in the vicinity of the particle. These methods are flexible and adaptable, and we apply them to a number of different situations. We reproduce old results from the literature, providing additional information and extensions of these results, correct some erroneous results, and consider situations that have not yet been examined. 

We begin in Sec.~\ref{sec:weyl} with a general description of the method, in the specific context of an electric charge held at rest in the spacetime of a nonrotating black hole. In Sec.~\ref{sec:background} we cast the background Schwarzschild metric in the Weyl form of Eq.~(\ref{weyl-intro}), and review some of its properties. In Sec.~\ref{sec:electric-charge} we place a charged particle in the spacetime, construct its vector potential and metric perturbation, and calculate the string's tension $T$ from $\gamma^{\rm axis}$. We obtain 
\begin{equation} 
T = \frac{mM}{r_0^2 (1-2M/r_0)^{1/2}} - \frac{2m^2 M^2}{r_0^4(1-2M/r_0)} - \frac{q^2 M}{r_0^3}
+ O(m^3, m q^2),  
\label{tension-em} 
\end{equation} 
where $M$ is the mass of the black hole, $m$ the particle's mass, $q$ its charge, and $r_0$ its position. The first term on the right of Eq.~(\ref{tension-em}) is recognized as $m a$, the particle's mass times its acceleration in the background Schwarzschild spacetime. The second term is a correction to this expression, which can be attributed to the particle's gravitational self-force; this contribution is negative, which reveals that the gravitational self-force is repulsive. The last term is the Smith-Will electromagnetic self-force, which is also repulsive. The error term in Eq.~(\ref{tension-em}) indicates that the tension is calculated through second order in both $m$ and $q$; the error is of third order. 

In the developments of Sec.~\ref{sec:electric-charge} we go at length to ensure that all quantities that appear in Eq.~(\ref{tension-em}) can be given an operational definition. The black-hole mass $M$ is thus identified with the Smarr mass \cite{smarr:73}, which can be defined in terms of geometric quantities on the deformed event horizon. The particle's mass $m$ is defined in terms of its energy-momentum tensor. The charge $q$ is similarly defined in terms of the current-density vector, but it can also be defined by Gauss' law applied to a closed surface surrounding the charge. The most difficult quantity to interpret is $r_0$, the coordinate position of the charge. We relate it to a redshift factor $\gothf_{\rm reg}$, the ratio of energies for a photon emitted at the particle and received at infinity; this is regularized by subtracting the (infinite) redshift contributed by the particle's local gravitational field. 

We continue in Sec.~\ref{sec:electric-dipole} with a calculation of the string tension produced by electric and magnetic dipoles in the spacetime of a nonrotating black hole. The self-force acting on dipoles was previously computed by L\'eaut\'e and Linet \cite{leaute-linet:84}, based on the expectation that only the regular part of the (electric or magnetic) field should be exerting this force. The field, however, is more singular for a dipole than for a point charge, and regularization may not be as straightforward as what was attempted by L\'eaut\'e and Linet. We find that indeed, our results differ from theirs. In Sec.~\ref{sec:scalar-charge} we replace the electric charge of Sec.~\ref{sec:electric-charge} with a scalar charge, and reproduce Wiseman's result \cite{wiseman:00} that the self-force vanishes; the tension is given by Eq.~(\ref{tension-em}) with the last term deleted. In an interesting extension of Wiseman's treatment, we find that the black hole and scalar charge system can be described by an {\it exact solution} to the Einstein-scalar equations. Finally, we show that the self-force on a scalar dipole is also zero.         

In Sec.~\ref{sec:extended}, the last section of the paper, we generalize our methods so that they can be applied to extended objects instead of point particles. We first calculate the string's tension for a system of two nonrotating black holes, and then consider a system of two massive, scalarized objects described by the Janis-Newman-Winicour metric \cite{janis-newman-winicour:68}. In the Appendix we provide additional insights into the spacetime of a Schwarzschild black hole perturbed by an electric charge, as described in Sec.~\ref{sec:electric-charge}.  

We hope that this survey of what our methods can achieve will convince the reader of their power, and that this reader will feel inspired to continue their exploration. In our view, the approach to the self-force provided by the Weyl class of metrics is a most compelling one. First, it provides a complete physical picture in which the external agent holding the particle is precisely identified as a massless string. Second, and more importantly, it permits a calculation of the force in which there is no need to renormalize a singular field. The approach, however, is limited by the restrictions inherent to the Weyl class of metrics: the spacetime must be static and axially symmetric, and the energy-momentum tensor must be such that $T^\rho_{\ z} = -T^z_{\ z}$. Fortunately, this last condition is fairly accommodating, being met by electromagnetic and massless scalar fields, and by point particles. With some work it should be possible to go beyond this class of spacetimes. For example, the restriction on the energy-momentum tensor can be lifted by incorporating a third gravitational potential in the metric, and a third potential can also allow the spacetime to become stationary (instead of merely static; see, for example, Ch.~13 of Ref.~\cite{griffiths-podolsky:09}). We shall leave these considerations for future work.      

\section{General scheme and strategy} 
\label{sec:weyl} 

We present our calculational scheme in the specific context of a point electric charge held at rest in the spacetime of a nonrotating black hole. The method is easily adaptable, and it will also be applied to an electric dipole, a magnetic dipole, a scalar charge, a scalar dipole, and extended objects. But for the time being we consider a point particle of mass $m$ and electric charge $q$ held in place outside a black hole of mass $M$. It is assumed that $m$ and $q$ are both much smaller than $M$, and we take $m$ and $q$ to be of the same order of magnitude. (There is no obstacle to letting $m$ be much smaller than $q$, or $q$ be much smaller than $m$.) We wish to calculate the force required to keep the particle in its place.  

\subsection{Metric, vector potential, and field equations} 

The spacetime is static and axially symmetric about the straight line that joins the particle to the black hole. The geometry of this spacetime is described by the Weyl metric of Eq.~(\ref{weyl-intro}), in which $U$ and $\gamma$ are functions of $\rho$ and $z$. The electric field produced by the particle is described by the vector potential
\begin{equation} 
A_\alpha = -\Phi\, \partial_\alpha t,
\end{equation} 
in which $\Phi$ is also a function of $\rho$ and $z$. The particle is placed on the $z$-axis, at $\rho = 0$ and $z = b$. 

The charged particle comes with a current-density vector 
\begin{equation} 
j^\alpha = q \int u^\alpha\, \frac{\delta\bigl(x-X(\tau)\bigr)}{\sqrt{-g}}\, d\tau 
\label{current}
\end{equation} 
and an energy-momentum tensor  
\begin{equation} 
T^{\alpha\beta} = m \int u^\alpha u^\beta\, \frac{\delta\bigl(x-X(\tau)\bigr)}{\sqrt{-g}}\, d\tau.  
\label{Tpart1} 
\end{equation} 
Here, $x^\alpha$ represents the coordinates of a spacetime event, $X^\alpha(\tau)$ describes the particle's world line, which is parametrized with proper time $\tau$, $u^\alpha = dX^\alpha/d\tau$ is the velocity vector, $\delta(x-X)$ is a four-dimensional delta function, and $g$ is the metric determinant. In the case of a static particle placed on the $z$-axis, the only nonvanishing components are
\begin{equation}
j^t = q e^{-2(U+\gamma)}\, \delta(\bm{x}-\bm{b}), \qquad
T^{tt} = m e^{-(U+2\gamma)} \delta(\bm{x}-\bm{b}),
\label{Tpart2} 
\end{equation}
in which $\bm{x}$ designates a spatial point with coordinates $(\rho,z,\phi)$, $\bm{b}$ denotes the position of the particle at $\rho=0$, $z=b$, and $\delta(\bm{x}-\bm{b}) := \rho^{-1} \delta(\rho) \delta(z-b) \delta(\phi)$ is a three-dimensional delta function. 

The field equations consist of Maxwell's equations 
\begin{equation} 
\nabla_\beta F^{\alpha\beta} = 4\pi j^\alpha, 
\end{equation} 
where $F_{\alpha\beta} = \nabla_\alpha A_\beta - \nabla_\beta A_\alpha$ is the electromagnetic field tensor, and Einstein's equations, 
\begin{equation} 
G^\alpha_{\ \beta} - 2F^{\alpha}_{\ \mu} F_{\beta}^{\ \mu} 
+ \frac{1}{2} \delta^{\alpha}_{\ \beta} F_{\mu\nu} F^{\mu\nu} 
= 8\pi T^{\alpha}_{\ \beta}, 
\end{equation} 
where $G^{\alpha}_{\ \beta}$ is the Einstein tensor. The $t$-component of Maxwell's equations --- the only nonvanishing one --- reduces to 
\begin{equation} 
\nabla^2 \Phi + 2 \bm{\nabla} U \cdot \bm{\nabla} \Phi = -4\pi q e^{-2U}\, \delta(\bm{x}-\bm{b}), 
\label{maxwell} 
\end{equation} 
where $\nabla^2 := \partial_{\rho\rho} + \rho^{-1} \partial_\rho + \partial_{zz}$ is the flat-space Laplacian operator in cylindrical coordinates, and $\bm{\nabla} \varphi \cdot \bm{\nabla} \psi := (\partial_\rho \varphi) (\partial_\rho \psi) + (\partial_z \varphi) (\partial_z \psi)$ for any two functions $\varphi$ and $\psi$ of $\rho$ and $z$. The $tt$- and $\phi\phi$- components of the Einstein field equations, one subtracted from the other, yield  
\begin{equation} 
\nabla^2 U + e^{2U} \bm{\nabla} \Phi \cdot \bm{\nabla} \Phi = -4\pi m e^{-U}\, \delta(\bm{x}-\bm{b}). 
\label{einsteinU} 
\end{equation} 
And the $\rho\rho$- and $zz$- components of the Einstein equations produce 
\begin{subequations} 
\label{einsteingamma} 
\begin{align} 
\frac{1}{\rho} \partial_\rho \gamma &= (\partial_\rho U)^2 - (\partial_z U)^2 
- e^{2U} \bigl[ (\partial_\rho \Phi)^2 - (\partial_z \Phi)^2 \bigr], \\ 
\frac{1}{\rho} \partial_z \gamma &= 2 (\partial_\rho U) (\partial_z U) 
- 2 e^{2U} (\partial_\rho \Phi) (\partial_z \Phi), 
\label{dz_gamma} 
\end{align} 
\end{subequations}  
respectively. We observe that Eqs.~(\ref{maxwell}) and (\ref{einsteinU}) for $\Phi$ and $U$ feature a distributional source term on the right-hand side, but that no such terms appear in Eqs.~(\ref{einsteingamma}) for $\gamma$. We observe also that Eqs.~(\ref{maxwell}) and (\ref{einsteinU}) do not involve $\gamma$; these equations are integrated first for $\Phi$ and $U$, and the results are inserted within Eqs.~(\ref{einsteingamma}) to obtain $\gamma$. 

\subsection{Perturbative expansion} 

The particle creates a perturbation of the Schwarzschild solution that describes the unperturbed black hole. To obtain the potentials in a perturbative series, we expand $U$ and $\Phi$ in powers of $\epsilon$, a book-keeping parameter (eventually set equal to one) that keeps track of the powers of $m$ and $q$. We write 
\begin{equation} 
U = U_0 + \epsilon U_1 + \epsilon^2 U_2 + O(\epsilon^3), \qquad 
\Phi = \epsilon \Phi_1 + \epsilon^2 \Phi_2 + O(\epsilon^3),  
\end{equation}  
in which $U_0$ describes a Schwarzschild black hole, and $U_n$, $\Phi_n$ (with $n=1,2$) are the perturbations created by the particle. Inserting the expansions within the field equations, we obtain the sequence of equations 
\begin{subequations} 
\label{sequence_U} 
\begin{align} 
\nabla^2 U_0 &= 0, \\ 
\nabla^2 U_1 &= -4\pi m e^{-U_0}\, \delta(\bm{x}-\bm{b}), 
\label{sequence_U1} \\ 
\nabla^2 U_2 + e^{2U_0} \bm{\nabla} \Phi_1 \cdot \bm{\nabla} \Phi_1
&= 4\pi m e^{-U_0} U_1\, \delta(\bm{x}-\bm{b})
\label{sequence_U2} 
\end{align} 
\end{subequations} 
for the gravitational potential, and the sequence 
\begin{subequations} 
\label{sequence_Phi} 
\begin{align} 
\nabla^2 \Phi_1 + 2 \bm{\nabla} U_0 \cdot \bm{\nabla} \Phi_1
&= -4\pi q e^{-2U_0}\, \delta(\bm{x}-\bm{b}),
\label{sequence_Phi1} \\ 
\nabla^2 \Phi_2 + 2 \bigl( \bm{\nabla} U_0 \cdot \bm{\nabla} \Phi_2 
+ \bm{\nabla} U_1 \cdot \bm{\nabla} \Phi_1 \bigr)
&= 8\pi q e^{-2U_0} U_1\, \delta(\bm{x}-\bm{b})
\label{sequence_Phi2} 
\end{align} 
\end{subequations} 
for the electrostatic potential. The sequences could be extended to higher orders in $\epsilon$, but we shall be satisfied with a truncation through order $\epsilon^2$. An issue that arises in the integration of the field equations is that $U_1$ is infinite at $\bm{x}=\bm{b}$, with the consequence that the source terms for $U_2$ and $\Phi_2$ do not make sense as distributions. As we shall see, we shall be able to evade this difficulty.

\subsection{String tension}

Once Eqs.~(\ref{sequence_U}) and (\ref{sequence_Phi}) have been integrated to reveal the potentials $U$ and $\Phi$ through order $\epsilon^2$, the results are inserted within Eqs.~(\ref{einsteingamma}) to obtain $\gamma$, also expanded through order $\epsilon^2$. Equation (\ref{dz_gamma}) implies that $\partial_z \gamma = 0$ when $\rho = 0$, except when the right-hand side of the equation is singular. It follows that $\gamma(\rho=0,z)$ is constant on any nonsingular portion of the axis, but the value of $\gamma(\rho=0,z)$ can jump from one constant to another when a singularity is encountered. Because $U$ and $\Phi$ are singular at $z=b$, where the particle is situated, we have that $\gamma(\rho=0,z)$ must jump at $z=b$. While $\gamma(\rho=0,z)$ can be taken to vanish\footnote{The opposite scenario is also possible: we can choose $\gamma(\rho=0,z>b)=0$, and find that $\gamma(\rho=0,z<b)$ cannot be zero. In this scenario the particle would be held in place by a strut situated between the black hole and the particle.}  for $z < b$ (between the black hole and the particle), it cannot be zero for $z > b$ (above the particle); we must have instead $\gamma^{\rm axis}:= \gamma(\rho=0,z>b) \neq 0$. 

A regular metric would have $\gamma$ vanish everywhere on the $z$-axis; a nonzero $\gamma^{\rm axis}$ reveals instead the presence of an angular deficit in the spacetime. The conical singularity, in turn, signals the presence of a material source on the axis, a string. Because a constant $\gamma^{\rm axis}$ implies a constant tension $T$, and because this is possible only if $T=\mu$, where $\mu$ is the string's mass density, the string is identified as a Nambu-Goto string. We recall that the tension is given by Eq.~(\ref{tension-intro}), and that it is directly proportional to the angular deficit. The sole gravitational manifestation of the string is this angular deficit; in particular, the string possesses a vanishing gravitational mass.  

The picture that emerges is that of a charged particle held in place at $z=b$ by being attached to a Nambu-Goto string, which extends from the particle out to infinity. The force required to keep the particle from falling toward the black hole, exerted by an observer holding the string at infinity, is equal to the string's tension. A calculation of $\gamma^{\rm axis}$, therefore, reveals the force acting on the particle.   

\subsection{Calculational scheme}
\label{subsec:scheme} 

We rely on Eqs.~(\ref{einsteingamma}) to calculate $\gamma(\rho=0,z>b)$. It is sufficient to integrate these equations in a small neighborhood of $z = b$, and for this purpose it is sufficient to know $U$ and $\Phi$ near $\rho=0$ and $z=b$. This observation defines our calculational strategy: Obtain the potentials locally, use this information to calculate the jump of $\gamma$ across $z = b$, and deduce the string's tension from $\gamma$.  

In most of the cases that we shall examine below, it is possible to obtain global solutions for $U_1$ and $\Phi_1$. We shall insert these in Eqs.~(\ref{sequence_U2}) and (\ref{sequence_Phi2}) to obtain $U_2$ and $\Phi_2$ near $\rho=0$ and $z=b$, and we shall then involve these local solutions in a calculation of $\gamma(\rho=0,z>b)$. The local analysis, however, does not return unique solutions for $U_2$ and $\Phi_2$, because it does not provide access to the required boundary conditions, either at infinity or at the black-hole horizon. The solutions, therefore, can only be obtained up to a number of unknown constants. It is a very fortunate circumstance that the calculation of $\gamma(\rho=0,z>b)$ is insensitive to the value of these constants.

In the cases to be considered in Sec.~\ref{sec:extended}, featuring extended objects instead of point particles, we shall have access to exact solutions for $U$ and $\Phi$. We shall nevertheless base the calculation of $\gamma$ on local expressions that are valid close to the extended objects. 

\section{Background spacetime}
\label{sec:background}

We begin in this section with a description of the background Schwarzschild metric, expressed in the Weyl coordinates $(t,\rho,z,\phi)$. The metric takes the form of Eq.~(\ref{weyl-intro}), and the Schwarzschild solution is given by $U = U_0$ and $\gamma = \gamma_0$, with 
\begin{equation}
e^{-2U_0} = \frac{R_+ + R_- - 2M}{R_+ + R_- + 2M}, \qquad
e^{2\gamma_0} = \frac{ (R_+ + R_-)^2 - 4M^2}{4 R_+ R_-},
\label{sch_metric} 
\end{equation}   
where
\begin{equation}
R_\pm := \sqrt{ \rho^2 + (z \pm M)^2 }.
\label{Rpm_def}
\end{equation}
The potential $U_0$, interpreted in Newtonian terms, is that of a rod of mass $M$ and length $2M$, with constant linear mass density $M/(2M) = 1/2$. In the Weyl coordinates, the event horizon is described by $\rho = 0$, $z = \pm M$.  

A test particle of mass $m$, at position $\rho = 0$ and $z = b$ in the Schwarzschild spacetime, possesses a velocity vector given by 
\begin{equation}
u^\alpha = \sqrt{\frac{b+M}{b-M}}\, t^\alpha,
\end{equation}
where $t^\alpha := (1,0,0,0)$ is the timelike Killing vector; the square-root factor is $e^{U_0}$ evaluated at $\rho=0$ and $z=b$. The particle's Killing energy is $E = -m u_\alpha t^\alpha$, or
\begin{equation}
E =m \sqrt{\frac{b-M}{b+M}}.
\label{killing_energy}
\end{equation}
The particle's acceleration vector is $a^\alpha := u^\beta \nabla_\beta u^\alpha$, and its only nonvanishing component is $a^z > 0$; its covariant magnitude $a := (g_{\alpha\beta} a^\alpha a^\beta)^{1/2}$ evaluates to
\begin{equation}
a = \frac{M}{(b-M)^{1/2} (b+M)^{3/2}}.
\label{acceleration}
\end{equation}

The transformation
\begin{equation}
\rho = r\sqrt{f}\, \sin\theta, \qquad
z = (r-M) \cos\theta, 
\end{equation}
where $f := 1-2M/r$, brings the metric to its usual Schwarzschild form. We have that $R_\pm = r - M \pm M\cos\theta$,
\begin{equation}
e^{-2U_0} = f, \qquad e^{2\gamma_0} = \frac{r^2 f}{R_+ R_-},
\end{equation}
and $d\rho^2 + dz^2 = (R_+ R_-/r^2)[ f^{-1}\, dr^2 + r^2\, d\theta^2 ]$. With this the metric turns into the familiar form  
\begin{equation}
ds^2 = -f\, dt^2 + f^{-1}\, dr^2 + r^2(d\theta^2 + \sin^2\theta\, d\phi^2).
\end{equation}
In terms of the Schwarzschild coordinates, the particle is at $r = r_0 := b + M$ and $\theta = 0$. Its Killing energy and acceleration are given by
\begin{equation}
E = m \sqrt{f_0}, \qquad  
a = \frac{M}{r_0^2 \sqrt{f_0}},
\label{accelerationS} 
\end{equation}
respectively, where $f_0 := 1-2M/r_0$. 

\section{Electric charge} 
\label{sec:electric-charge} 

Next we calculate the perturbations $U_1$, $\Phi_1$, $U_2$, and $\Phi_2$ associated with a point electric charge, obtain $\gamma$ on the axis, and calculate the string's tension. We follow the strategy outlined in Sec.~\ref{sec:weyl}, and provide the missing details. 

\subsection{First-order perturbation} 
\label{sec:first-order}

We begin by constructing the first-order corrections to the metric and vector potential created by the charged particle at $\rho = 0$, $z=b$. We write $U = U_0 + \epsilon U_1 + O(\epsilon^2)$, $\Phi = \epsilon \Phi_1 + O(\epsilon^2)$, and $\gamma = \gamma_0 + \epsilon \gamma_1 + O(\epsilon^2)$, and work to first order in $\epsilon$. 

\subsubsection{Potentials} 

The gravitational potential $U_1$ must be a solution to Eq.~(\ref{sequence_U1}). It is easy to see that it is given by  
\begin{equation} 
U_1 = \frac{E}{D} = m \sqrt{\frac{b-M}{b+M}} \frac{1}{D},
\label{U1} 
\end{equation} 
where $E$ is the Killing energy of Eq.~(\ref{killing_energy}), and
\begin{equation}
D := \sqrt{\rho^2 + (z-b)^2}
\end{equation}
is the Euclidean distance to the particle.  This potential can also be viewed as the linearized approximation to the Schwarzschild potential of a black hole of mass $E$ situated at $z = b$. It is worth noting that in this view, the error introduced in the linearization is of order $E^3$, and not $E^2$ as might be expected. The potential can also be viewed as the exact representation of a Curzon-Chazy particle \cite{curzon:24, chazy:24}. 

The electrostatic potential $\Phi_1$ must be a solution to Eq.~(\ref{sequence_Phi1}). Such a potential was constructed by Copson \cite{copson:28} and then corrected by Linet \cite{linet:76}. In the Weyl coordinates it is given by   
\begin{equation} 
\Phi_1 = \frac{q}{(b+M)(R_+ + R_- + 2M)} \biggl[ 
\frac{(b+M) R_{-} + (b-M) R_{+}}{D} + 2M \biggr]. 
\label{Phi_CL} 
\end{equation} 
An application of Gauss' law confirms that a small sphere surrounding $(\rho,z) = (0,b)$ contains a charge $q$, and that the total charge in the spacetime is also $q$. The second property was not verified in Copson's original solution; Linet added the $2M$ term within the square brackets in Eq.~(\ref{Phi_CL}), a monopole solution to Maxwell's equation, to restore the correct value for the total charge.

Next we insert the expansions $U = U_0 + \epsilon U_1 + O(\epsilon^2)$, $\Phi = \epsilon \Phi_1 + O(\epsilon^2)$ within Eqs.~(\ref{einsteingamma}) and determine $\gamma_1$. We find that the general solution to these equations features an integration constant, and we choose this constant so that $\gamma_1(\rho=0,z) = 0$ when $M < z < b$, that is, between the black hole and the particle. With this choice we find that
\begin{equation} 
\gamma_1 = \frac{m}{(b-M)^{1/2}(b+M)^{3/2}} \biggl[ 
\frac{(b+M) R_- - (b-M)R_+}{D} + 2M \biggr].
\label{gamma1}
\end{equation}
With this solution we find that $\gamma_1(\rho=0,z) = 4ma$ when $z > b$ (above the particle), with $a$ denoting the acceleration of Eq.~(\ref{acceleration}). We also have that $\gamma_1(\rho=0,z) = 4ma$ when $z < -M$ (on the other side of the black hole). According to this and Eq.~(\ref{tension-intro}), the particle is held in place with the help of a Nambu-Goto string with tension
\begin{equation}
T = \epsilon\, m a + O(\epsilon^2).
\label{tension_epsilon} 
\end{equation}
A string is also attached to the black hole, extending from $z = -M$ to $z = -\infty$. The tension in this string is also equal to $\epsilon m a + O(\epsilon^2)$. In a beautiful illustration of Newton's second and third laws, applied to a fully relativistic situation, the force required to keep the particle from falling toward the black hole is equal to the force required to keep the black hole from falling toward the particle, and each force is equal to $m a$, the particle's mass times its acceleration. 

In the usual Schwarzschild coordinates we have that
\begin{subequations}
\begin{align}
U_1 &= \frac{m\sqrt{f_0}}{D}, \\
\Phi_1 &= \frac{q}{r_0 r} \biggl[ \frac{(r-M)(r_0-M) - M^2\cos\theta}{D} + M \biggr],\\
\gamma_1 &= \frac{2mM}{r_0^2\sqrt{f_0}} \biggl[ \frac{r-M - (r_0-M) \cos\theta}{D} + 1 \biggr],
\label{g1_Schw} 
\end{align}
\end{subequations}
where $f_0 := 1-2M/r_0$ and $D$ is now given by 
\begin{equation}
D = \bigl[ (r-M)^2 - 2(r-M)(r_0-M)\cos\theta + (r_0-M)^2 - M^2\sin^2\theta \bigr]^{1/2}.
\label{D_Schw} 
\end{equation}
In this description, $\gamma_1(r,\theta=0)$ is zero when $2M < r < r_0$, and equal to $4ma$ when $r > r_0$, with $a$ now given by Eq.~(\ref{accelerationS}); we also have that $\gamma_1(r,\theta=\pi) = 4ma$. 
At first order in $\epsilon$, the metric of the perturbed Schwarzschild black hole is given by
\begin{equation}
ds^2 = -(1-2\epsilon U_1) f\, dt^2
+ \bigl[1 + 2\epsilon(U_1 + \gamma_1) \bigr] \bigl( f^{-1}\, dr^2 + r^2\, d\theta^2 \bigr)
+ (1+2\epsilon U_1) r^2 \sin^2\theta\, d\phi^2,
\label{metricpert} 
\end{equation}
and the vector potential is $A_\alpha = -\epsilon \Phi_1\, \partial_\alpha t$.

\subsubsection{Black-hole properties} 

The metric of Eq.~(\ref{metricpert}) can be used to calculate how the black hole is affected by the perturbation. It is evident that the event horizon continues to be situated at $r = 2M$, where $g_{tt} = 0$, and that the induced metric on the horizon is given by the $\theta$-$\theta$ and $\phi$-$\phi$ components of the metric, with $U_1$ and $\gamma_1$ evaluated at $r=2M$. These horizon values are
\begin{equation}
U_1(r=2M) = \frac{m\sqrt{f_0}}{r_0 - M - M\cos\theta}, \qquad
\gamma_1(r=2M) = \frac{2mM}{r_0 \sqrt{f_0}} \frac{1-\cos\theta}{r_0-M-M\cos\theta},
\label{horizon_values} 
\end{equation}
and we also have that $\Phi_1(r=2M) = q/r_0$. The black-hole area is calculated to be
\begin{equation}
A = 16\pi M^2 \biggl( 1 + \epsilon \frac{2m}{r_0\sqrt{f_0}} \biggr) + O(\epsilon^2)
= 16\pi M^2 \biggl( 1 + \epsilon \frac{2m}{\sqrt{b^2-M^2}} \biggr) + O(\epsilon^2).
\label{horizon_area} 
\end{equation} 
The surface gravity is obtained from $\kappa^2 = -\frac{1}{2} (\nabla_\alpha t_\beta) (\nabla^\alpha t^\beta)$, where the right-hand side is evaluated on the horizon. We find
\begin{equation}
\kappa = \frac{1}{4M} \biggl( 1 - \epsilon \frac{2m}{r_0\sqrt{f_0}} \biggr) + O(\epsilon^2)
= \frac{1}{4M} \biggl( 1 - \epsilon \frac{2m}{\sqrt{b^2-M^2}} \biggr) + O(\epsilon^2), 
\label{surface_gravity}
\end{equation}
and observe that in accordance with the zeroth-law of black-hole mechanics, the surface gravity is constant on the horizon.

It follows from Eqs.~(\ref{horizon_area}) and (\ref{surface_gravity}) that the Smarr mass of the black hole, defined by \cite{smarr:73} $M_{\rm Smarr} := \kappa A/(4\pi)$, is given by
\begin{equation}
M_{\rm Smarr} = M + O(\epsilon^2). 
\label{Smarr_mass} 
\end{equation}
The mass parameter $M$ can therefore be related to geometric objects defined on the perturbed event horizon. It is possible to formulate a first law of black-hole mechanics for the deformed black hole. For a quasi-static process in which $m$ is kept fixed (but $M$ and $r_0$ are allowed to vary), it takes the form of
\begin{equation}
d M_{\rm tot} = \frac{\kappa}{8\pi}\, dA - \lambda\, dT,
\label{first-law}
\end{equation} 
where $M_{\rm tot} := M + E = M + \epsilon\, m\sqrt{f_0} + O(\epsilon^2)$ is the total mass in the spacetime, $T$ is the string's tension of Eq.~(\ref{tension_epsilon}), and
\begin{equation}
\lambda := r_0 + O(\epsilon) 
\label{thermolength} 
\end{equation}
is the string's ``thermodynamic length'', a quantity defined by the first law itself. It is noteworthy that $M_{\rm tot}$ plays the role of enthalpy (instead of energy) in Eq.~(\ref{first-law}). Appels, Gregory, and Kubiz\v{n}\'ak \cite{appels-gregory-kubiznak:17} have shown that an enthalpy formulation of the first law is what should be expected of black holes with angular deficits.  

\subsubsection{Regularized redshift}

We consider, in a spacetime with the metric of Eq.~(\ref{weyl-intro}), a photon emitted at $z = b$ and received at $z = \infty$; the photon is assumed to travel on the $z$-axis, with $\rho = 0$. The photon's energy at the emission event, as measured by a static observer at $z = b$, is denoted $E(b)$, while its energy at reception, measured by a second static observer at infinity, is denoted $E(\infty)$. The energies are related by the redshift formula
\begin{equation}
E(\infty) = \gothf\, E(b), \qquad \gothf = e^{-U(b)},
\end{equation}
where $U(b)$ is the gravitational potential evaluated at $\rho = 0$, $z = b$.

With $U = U_0 + \epsilon U_1 + O(\epsilon^2)$ we find that $\gothf$ is ill-defined, because $U_1$ is formally infinite at $\rho = 0$, $z = b$. In the spirit of Detweiler's redshift invariant \cite{detweiler:08}, we regularize $\gothf$ by removing $U_1$ from the accounting of the gravitational potential. Operationally, this amounts to letting the photon be emitted slightly away from the particle, and subtracting from the overall redshift --- now finite --- the piece contributed by the particle's local gravitational field, described by $U_1$. The regularized redshift is then
\begin{equation}
\gothf_{\rm reg} = e^{-U_0(b)} + O(\epsilon^2)
= \sqrt{\frac{b-M}{b+M}} + O(\epsilon^2)
= \sqrt{f_0} + O(\epsilon^2).
\label{redshift1} 
\end{equation}
This equation can be inverted to express $b$ and $r_0$ in terms of the regularized redshift. We have that
\begin{equation}
b/M = \frac{1 + \gothf_{\rm reg}^2}{1 - \gothf_{\rm reg}^2} + O(\epsilon^2), \qquad
r_0/M = \frac{2}{1 - \gothf_{\rm reg}^2} + O(\epsilon^2).   
\label{redshift2}
\end{equation}
In this way, the coordinate position of the particle can be expressed in  terms of a meaningful observable.   

The prescription of Eq.~(\ref{redshift1}) can be related to a standard regularization procedure of post-Newtonian theory, in which a formally infinite quantity is replaced by its {\it Hadamard partie finie} \cite{blanchet-faye:00}. To define this, we introduce local polar coordinates $(s,\alpha)$ near $\rho=0, z=b$, given by $\rho = s \sin\alpha$, $z = b + s \cos\alpha$, and we consider a function $h(s,\alpha)$ that is singular in the limit $s \to 0$. More precisely, we assume that $h$ admits a Laurent series of the form $h(s,\alpha) = \sum_{n \geq -N} h_n(\alpha) s^n$ near $s=0$; the series is taken to begin at order $s^{-N}$, with $N > 0$. Then its {\it Hadamard partie finie} is defined to be
\begin{equation}
\lfloor h \rfloor := \frac{1}{2} \int_0^\pi h_0(\alpha) \sin\alpha\, d\alpha; 
\end{equation}
it is the average over all angles of the zeroth-order term in the series. In our application, $h = \gothf = [1 - \epsilon E/s + O(\epsilon^2)] e^{-U_0}$, with
\begin{equation}
e^{-U_0} = \sqrt{\frac{b-M}{b+M}} - \frac{M s \cos\alpha}{(b-M)^{1/2}(b+M)^{3/2}} + O(s^2).
\end{equation}
Extracting the $s=0$ term in the Laurent series and integrating over $\alpha$, we arrive at $\lfloor \gothf \rfloor = (b-M)^{1/2}/(b+M)^{1/2} + O(\epsilon^2)$, in agreement with Eq.~(\ref{redshift1}). Hadamard's regularization procedure will be exploited again in the following subsection. 

\subsection{Second-order perturbation}
\label{sec:second-order} 

We proceed to the next order in the perturbative expansion, write $U = U_0 + \epsilon U_1+ \epsilon^2 U_2 + O(\epsilon^3)$, $\Phi = \epsilon \Phi_1 + \epsilon^2 \Phi_2 + O(\epsilon^3)$, and work to second order in $\epsilon$. The goal is to obtain $U_2$, $\Phi_2$ by integrating Eqs.~(\ref{sequence_U2}) and (\ref{sequence_Phi2}), respectively. Unfortunately these equations cannot be solved exactly, but as was stated in Sec.~\ref{subsec:scheme}, it is sufficient for our purposes to obtain $U_2$ and $\Phi_2$ in a small neighborhood around the particle. 

To achieve this it is helpful to reformulate the field equations in terms of the local polar coordinates $(s, \alpha)$ introduced previously. These are defined by $\rho = s \sin\alpha$, $z = b + s \cos\alpha$, and with them the metric becomes 
\begin{equation}
ds^2 = -e^{-2U}\, dt^2 + e^{2(U+\gamma)} \bigl( ds^2 + s^2\, d\alpha^2 \bigr) + e^{2U} s^2 \sin^2\alpha\, d\phi^2, 
\label{metricBH_local} 
\end{equation} 
The previously calculated potentials take the local form 
\begin{subequations} 
\label{01pot_local} 
\begin{align} 
e^{-2U_0} &= \frac{b-M}{b+M} + \frac{2M \cos\alpha}{(b+M)^2}\, s 
+ \frac{M\bigl[ b - (3b-2M)\cos^2\alpha \bigr]}{(b-M)(b+M)^3}\, s^2 
+ O(s^3), \\ 
U_1 &= m \sqrt{\frac{b-M}{b+M}} \frac{1}{s}, \\ 
\Phi_1 &= \frac{b-M}{b+M} \frac{q}{s} + \frac{qM}{(b+M)^2} (1 + \cos\alpha) 
- \frac{qM}{2(b-M)(b+M)^3} \bigl[ (3b-M)\cos\alpha - (b+M) \bigr] (1 + \cos\alpha) s 
\nonumber \\ & \quad \mbox{} 
+ O(s^2). 
\end{align} 
\end{subequations} 
The differential operators that occur in Eqs.~(\ref{sequence_U}) and (\ref{sequence_Phi}) become
\begin{subequations}
\begin{align}
\nabla^2 \psi &= 
\frac{1}{s^2} \partial_{s} \bigl( s^2\, \partial_{s} \psi \bigr)
+ \frac{1}{s^2 \sin\alpha}\, \partial_{\alpha} \bigl( \sin\alpha\, \partial_\alpha \psi \bigr), \\
\bm{\nabla} \psi \cdot \bm{\nabla} \varphi &= (\partial_s \psi) (\partial_s \varphi)
+ \frac{1}{s^2} (\partial_\alpha \psi) (\partial_\alpha \varphi)
\end{align}
\end{subequations}
in the local polar coordinates; here $\psi$ and $\varphi$ are any functions of $s$ and $\alpha$. 

For the moment we ignore the distributional term on the right-hand side of Eq.~(\ref{sequence_U2}), and find a solution to the equation by making the ansatz
\begin{equation}
U_2 = \frac{u_{-2}}{s^2} + \frac{u_{-1}}{s} + u_0 + O(s),
\label{U2sol} 
\end{equation}
where the coefficients $u_j$ are functions of $\alpha$. These are determined by integrating Eq.~(\ref{sequence_U2}) order-by-order in $s$, and demanding that the solutions be regular at $\alpha = 0$ and $\alpha = \pi$. We obtain
\begin{subequations}
\begin{align}
u_{-2} &= -\frac{q^2(b-M)}{2 (b+M)}, \\
u_{-1} &= -\frac{q^2 M}{(b+M)^2} \cos\alpha, \\
u_0 &= \frac{q^2 M}{(b+M)^3} \cos\alpha + \frac{q^2 Mb}{(b-M)(b+M)^3} \cos^2\alpha + \mu_0,
\end{align}
\end{subequations}
where $\mu_0$ is an arbitrary constant.

A complete solution must also account for the source term in Eq.~(\ref{sequence_U2}). The delta function on the right-hand side calls for the inclusion of a term $\delta E/s$ in $U_2^{\rm part}$, with a shift in energy parameter formally given by $\delta E = -m e^{-U_0} U_1$, with the right-hand side evaluated at the particle's position. This quantity is actually infinite, but meaning can be given to it by replacing it by its {\it Hadamard partie finie}, as was done in the preceding subsection. The regularization prescribes $\delta E = -m \lfloor e^{-U_0} U_1 \rfloor$, and making use of Eqs.~(\ref{01pot_local}) to perform the calculation, we find that $\delta E = 0$.   

To the solution of Eq.~(\ref{U2sol}) we might have added any solution to the homogeneous version of Eq.~(\ref{sequence_U2}), in which we set both the distributional source term and $\Phi_1$ to zero, and thereby recover Laplace's equation. The constant $\mu_0$ reflects this freedom, but we might also have included multipolar terms of the form $s^{-(\ell+1)} P_\ell(\cos\alpha)$ with $\ell \geq 1$, where $P_\ell$ are Legendre polynomials. That such terms must be excluded can be justified on the grounds that Eq.~(\ref{sequence_U2}) does not feature matching distributional sources (involving derivatives of delta functions) for higher multipoles. Additional evidence in favor of this exclusion comes from the asymptotic matching that we carry out in Appendix~\ref{sec:BHPartView}. The solution of Eq.~(\ref{U2sol}) is complete.  

To obtain a solution to Eq.~(\ref{sequence_Phi2}), first without the source term on the right-hand side, we write
\begin{equation}
\Phi_2 = \frac{p_{-2}}{s^2} + \frac{p_{-1}}{s} + p_0 + O(s),
\label{Phi2sol} 
\end{equation}
where each $p_j$ is a function of $\alpha$. Integrating Eq.~(\ref{sequence_Phi2}) order-by-order in $s$, we obtain
\begin{subequations}
\begin{align}
p_{-2} &= -mq \biggl( \frac{b-M}{b+M} \biggr)^{3/2}, \\ 
p_{-1} &= -2mq \frac{M (b-M)^{1/2}}{(b+M)^{5/2}} \cos\alpha, \\ 
p_0 &= mq \frac{M}{2(b-M)^{1/2}(b+M)^{7/2}} \bigl[ (5b-3M)\cos^2\alpha + 2(b-M) \cos\alpha \bigr]
+ \nu_0, 
\end{align}
\end{subequations}
where $\nu_0$ is an arbitrary constant. To account for the delta function on the right-hand side of Eq.~(\ref{sequence_Phi2}), we should insert a term $\sigma^2\delta q/s$ in $\Phi_2^{\rm part}$, where $\sigma^2 := (b-M)/(b+M)$, and where the shift in charge parameter is given by the regularized expression $\delta q = -m \sigma^{-2} \lfloor e^{-2U_0} U_1 \rfloor$. Performing the calculation of the {\it Hadamard partie finie} as we did previously, we find that $\delta q =0$. This conclusion is supported by an application of Gauss' law: For a small sphere surrounding the particle, an electrostatic potential that contains a term $\sigma^2 \delta q/s$ would produce an enclosed charge equal to $q + \delta q$; because the particle's charge is $q$, we must indeed set $\delta q = 0$. This corroboration lends considerable credence to the regularization procedure.      

To the solution of Eq.~(\ref{Phi2sol}) we might have added any solution to the homogeneous version of Eq.~(\ref{sequence_Phi2}), in which the distributional source term, $U_1$, and $\Phi_1$ are all set to zero. The constant $\nu_0$ reflects this freedom, and we again rule out singular terms corresponding to multipole moments, given the absence of matching distributional terms on the right-hand side, and on the basis of the asymptotic matching to be carried out in Appendix~\ref{sec:BHPartView}. 

Equations (\ref{01pot_local}), (\ref{U2sol}), and (\ref{Phi2sol}) provide us with local expressions for the potentials $U = U_0 + \epsilon U_1+ \epsilon^2 U_2 + O(\epsilon^3)$ and $\Phi = \epsilon \Phi_1 + \epsilon^2 \Phi_2 + O(\epsilon^3)$. The description of the black-hole view is as complete as we can make it. 

\subsection{String tension} 
\label{subsec:gamma} 

With $U$ and $\Phi$ now at hand, we are ready to tackle the calculation of $\gamma$ on the axis. As we have seen, the value of $\gamma(\rho=0,z)$ can jump from one constant to another at the singular point $z = b$. The values of $\gamma$ on each side of this singularity are related by  
\begin{equation} 
\gamma(z>b) - \gamma(z<b) = \int_C (\partial_\rho \gamma\, d\rho + \partial_z \gamma\, dz), 
\end{equation} 
where $C$ is any contour in the $\rho$-$z$ plane that links the points at $z>b$ and $z<b$. It is convenient to choose $C$ to be a half-circle described by $\rho = s \sin\alpha$, $z = b + s\cos\alpha$, where $s$ is constant and $\pi \geq \alpha \geq 0$. Choosing $\gamma(z<b) = 0$ --- no strut between black hole and particle --- we then find that $\gamma(z>b)$ is given by
\begin{equation}
\gamma^{\rm axis} = -\int_0^\pi \partial_\alpha \gamma\, d\alpha,
\label{gamma_axis}
\end{equation}
where 
\begin{align} 
\partial_{\alpha} \gamma &= -s^2 \sin\alpha\cos\alpha\, (\partial_s U)^2
     + 2 s \sin^2\alpha (\partial_s U)(\partial_\alpha U)
     + \sin\alpha\cos\alpha (\partial_\alpha U)^2
\nonumber \\ & \quad \mbox{}
     - e^{2U} \Bigl[ -s^2 \sin\alpha\cos\alpha\, (\partial_s \Phi)^2
     + 2 s \sin^2\alpha (\partial_s \Phi)(\partial_\alpha \Phi)
     + \sin\alpha\cos\alpha (\partial_\alpha\Phi)^2 \Bigr]
\label{dalpha_gamma} 
\end{align}
can be deduced from Eq.~(\ref{einsteingamma}). The coordinates $(s,\alpha)$ were introduced back in Sec.~\ref{sec:second-order}. 

We insert $U = U_0 + \epsilon U_1 + \epsilon^2 U_2 + O(\epsilon^3)$ and $\Phi = \epsilon \Phi_1 + O(\epsilon^2)$ within Eq.~(\ref{dalpha_gamma}), with $U_0$ given by Eq.~(\ref{01pot_local}), $U_1$ by Eq.~(\ref{U1}), $U_2$ by Eq.~(\ref{U2sol}), and $\Phi_1$ given by Eq.~(\ref{01pot_local}). We expand $\partial_\alpha \gamma$ through order $\epsilon^2$, and notice that $\Phi_2$ is not required in this calculation. We evaluate the integral of Eq.~(\ref{gamma_axis}), and observe that as it should, the outcome is independent of the contour radius $s$. We arrive at  
\begin{equation} 
\gamma^{\rm axis} = \epsilon\, \frac{4mM}{(b-M)^{1/2}(b+M)^{3/2}} - \epsilon^2\, \frac{4q^2M}{(b+M)^3}
+ O(\epsilon^3). 
\label{gamma_axis_final}  
\end{equation} 
It is interesting to note that the contribution proportional to $q^2$ in $\gamma^{\rm axis}$ comes entirely from the $\Phi$ terms in $\partial_\alpha \gamma$. The $q^2$ terms in $U$ combine with corresponding ones in $\Phi$ to cancel out contributions that would otherwise depend on $s$. It is also interesting to observe that while $\partial_\alpha \gamma$ features terms of order $m^2$, these do not survive the integration; $\gamma^{\rm axis}$ is therefore free of such terms. Finally, we point out that the calculation is completely insensitive to the value of the constants $\mu_0$ and $\mu_1$ that were introduced in Eq.~(\ref{U2sol}); these contributions also do not survive the integration over $\alpha$.  

We substitute Eq.~(\ref{gamma_axis_final}) into Eq.~(\ref{tension-intro}), expand in powers of $\epsilon$, and find that the string's tension is given by  
\begin{equation} 
T = \epsilon\, \frac{mM}{(b-M)^{1/2}(b+M)^{3/2}} - \epsilon^2\, \frac{2m^2 M^2}{(b-M)(b+M)^3}
- \epsilon^2\, \frac{q^2 M}{(b+M)^3} + O(\epsilon^3). 
\label{tension} 
\end{equation} 
The first term is recognized as $m a$, the particle's mass times its acceleration in the background Schwarzschild spacetime, as given by Eq.~(\ref{acceleration}). The second term is a gravitational self-force correction to this expression. The last term is the Smith-Will electromagnetic self-force. It is useful to recall that in the usual Schwarzschild coordinates, the particle is situated at $r_0 = b + M$; with this translation, we recover Eq.~(\ref{tension-em}). 

The quantities that appear in Eq.~(\ref{tension}) are all well-defined in operational terms. The particle's mass $m$ is defined in terms of the particle's energy-momentum tensor in Eq.~(\ref{Tpart1}), and its charge $q$ is defined in terms of the current density of Eq.~(\ref{current}); alternatively, the charge can be defined by Gauss' law applied to a small surface surrounding the particle. Up to terms of order $\epsilon^2$, the black-hole mass $M$ was identified with the Smarr mass, which is defined in terms of geometric quantities (surface gravity, area) on the horizon. And $b$, which designates the coordinate position of the particle, can be related by Eq.~(\ref{redshift2}) to the regularized redshift $\gothf_{\rm reg}$ of a photon emitted close to the particle and received at infinity.   

\section{Electric and magnetic dipoles}
\label{sec:electric-dipole}

In this section we examine the case of an electric dipole of mass $m$ and dipole moment $p$ held in place outside a nonrotating black hole of mass $M$. As before, the particle is placed at $r = r_0$ and $\theta = 0$, or at $\rho = 0$ and $z = b$, with $r_0 = b - M$. To respect the required axial symmetry, the dipole moment points in the radial direction, along the $z$-axis. We wish to calculate the force required to hold the dipole in place, and to achieve this we adapt the strategy described in Sec.~\ref{sec:weyl} to this new situation. Most of it is unchanged; the only difference concerns the source terms in Eqs.~(\ref{maxwell}) and (\ref{sequence_Phi}), which are now proportional to a $z$-derivative of $\delta(\bm{x}-\bm{b})$.  We calculate the perturbations $U_1$, $\Phi_1$, and $U_2$ associated with the point dipole (we omit $\Phi_2$, which is not needed), obtain $\gamma$ on the axis, and then calculate the string's tension. 

A duality transformation takes the field of an electric dipole to that of a magnetic dipole, and these field configurations come with the same distribution of energy-momentum tensor. The force on a magnetic dipole is therefore calculated in exactly the same way, and the calculation returns the same answer; this computation is detailed in Sec.~\ref{sec:magnetic-dipole}. 

\subsection{First-order perturbation} 
\label{sec:first-order-dipole}

We begin with the first-order corrections to the metric and vector potential created by the dipole at $\rho = 0$, $z=b$. We write $U = U_0 + \epsilon U_1 + O(\epsilon^2)$, $\Phi = \epsilon \Phi_1 + O(\epsilon^2)$, and $\gamma = \gamma_0 + \epsilon \gamma_1 + O(\epsilon^2)$, and work to first order in $\epsilon$. 

The gravitational potential $U_1$ is the same as in Sec.~\ref{sec:first-order}, 
\begin{equation} 
U_1 = m \sqrt{\frac{b-M}{b+M}} \frac{1}{D},
\label{U1-dipole} 
\end{equation} 
where $D := \sqrt{\rho^2 + (z-b)^2}$.

The electrostatic potential of a dipole can be obtained by superposing two monopole potentials, one for a negative charge $-q$ at $z = b - \delta b$, the other for a positive charge $+q$ at $z = b + \delta b$, and taking the limit $\delta b \to 0$ keeping $2q\delta b$ fixed. This gives
\begin{equation}
\Phi_1^{\rm dipole} = p \sqrt{\frac{b-M}{b+M}} \frac{\partial}{\partial b} \Phi_1^{\rm monopole},
\end{equation}
where $\Phi_1^{\rm monopole}$ is the Copson-Linet potential of Eq.~(\ref{Phi_CL}), and $p$ is a covariant dipole moment, which differs from $2q\delta b$ by a factor of $(b+M)^{1/2}/(b-M)^{1/2}$, the ratio of proper distance to coordinate distance along the $z$-axis. Performing the calculation yields
\begin{equation}
\Phi_1 = \frac{p (b-M)^{1/2}}{(b+M)^{3/2} (R_+ + R_- + 2M)} \biggl\{
\frac{(z-b) \bigl[ (b+M)R_- + (b-M) R_+ \bigr]}{D^3}
+ \frac{2M R_+}{(b+M) D} - \frac{2M}{b+M} \biggr\},
\label{Phi-dipole} 
\end{equation}
where $R_\pm$ was introduced in Eq.~(\ref{Rpm_def}), and where we henceforth omit the label ``dipole'' on the potential. In the usual Schwarzschild coordinates, the dipole potential is expressed as
\begin{equation}
\Phi_1 = \frac{p \sqrt{f_0}}{r_0 r} \biggl\{
\frac{ \bigl[(r-M)(r_0-M) - M^2\cos\theta \bigr]\bigl[ (r-M)\cos\theta - (r_0-M) \bigr] }{D^3}
+ \frac{M (r-M+M\cos\theta)}{r_0 D} - \frac{M}{r_0} \biggr\},
\label{Phi-dipole-Schw} 
\end{equation}
where $f_0 := 1-2M/b_0$, and $D$ is now given by Eq.~(\ref{D_Schw}). 

Because $\Phi_1$ does not enter in the calculation of $\gamma_1$, we find that the result of Eq.~(\ref{gamma1}) is unchanged. This implies that at first order in $\epsilon$, the tension is again given by Eq.~(\ref{tension_epsilon}).  And because the first-order metric is the same as in Sec.~\ref{sec:first-order}, the black-hole area and surface gravity are still given by Eqs.~(\ref{horizon_area}) and (\ref{surface_gravity}), respectively. The Smarr mass of the black hole is still identified with $M$, and the first law continues to be given by Eq.~(\ref{first-law}). Finally, the regularized redshift of a photon emitted close to the dipole and received at infinity is still given by Eq.~(\ref{redshift2}). 

\subsection{Second-order perturbation}
\label{sec:second-dipole} 

We next proceed to second order in the perturbative expansion. We write $U = U_0 + \epsilon U_1+ \epsilon^2 U_2 + O(\epsilon^3)$, $\Phi = \epsilon \Phi_1 + \epsilon^2 \Phi_2 + O(\epsilon^3)$, and obtain $U_2$ by integrating Eqs.~(\ref{sequence_U2}). We again rely on the local polar coordinates $(s,\alpha)$, and expand everything in powers of $s$. We require 
\begin{subequations} 
\label{01pot_dipole} 
\begin{align} 
e^{-2U_0} &= \frac{b-M}{b+M} + \frac{2M \cos\alpha}{(b+M)^2}\, s 
+ \frac{M}{(b-M)(b+M)^3} \Bigl[ b - (3b-2M)\cos^2\alpha \Bigr]\, s^2 
\nonumber \\ & \quad \mbox{} 
- \frac{M}{(b-M)^2(b+M)^4} \Bigl[ (3b^2-2Mb+M^2)\cos\alpha - (5b^2-6Mb+3M^2)\cos^3\alpha \Bigr]\, s^3
\nonumber \\ & \quad \mbox{} 
- \frac{M}{4(b-M)^3(b+M)^5} \Bigl[ b(3b^2-2Mb+3M^2)
- 2(15b^3-18Mb^2+19M^2b-4M^3)\cos^2\alpha
\nonumber \\ & \quad \mbox{} 
+ (35b^3-58Mb^2+59M^2b-16M^3)\cos^4\alpha \Bigr]\, s^4 
+ O(s^5), \\ 
U_1 &= m \sqrt{\frac{b-M}{b+M}} \frac{1}{s}, \\ 
\Phi_1 &= p \frac{(b-M)^{3/2}}{(b+M)^{3/2}} \frac{\cos\alpha}{s^2}
+ p M \frac{(b-M)^{1/2}}{(b+M)^{5/2}} \Bigl[ 1 + \cos^2\alpha \Bigl]\, \frac{1}{s}
\nonumber \\ & \quad \mbox{} 
- p M \biggl[ \frac{(b-M)^{1/2}}{(b+M)^{7/2}}
- \frac{1}{2(b-M)^{1/2}(b+M)^{5/2}}\, \cos\alpha 
+ \frac{3b-M}{2(b-M)^{1/2}(b+M)^{7/2}}\, \cos^3\alpha \biggr]
\nonumber \\ & \quad \mbox{} 
- p M \biggl[ \frac{M}{2(b-M)^{1/2}(b+M)^{9/2}}
- \frac{(b-M)^{1/2}}{(b+M)^{9/2}}\, \cos\alpha
+ \frac{3b^2+M^2}{2(b-M)^{3/2}(b+M)^{9/2}}\, \cos^2\alpha
\nonumber \\ & \quad \mbox{} 
- \frac{5b^2-3Mb+2M^2}{2(b-M)^{3/2}(b+M)^{9/2}}\, \cos^4\alpha \biggr]\, s
\nonumber \\ & \quad \mbox{} 
+ p M \biggl[ \frac{b}{2(b-M)^{1/2}(b+M)^{11/2}}
- \frac{3b^3-9Mb^2+17M^2b-3M^3}{8(b-M)^{5/2}(b+M)^{11/2}}\, \cos\alpha
- \frac{3b-2M}{2(b-M)^{1/2}(b+M)^{11/2}}\, \cos^2\alpha
\nonumber \\ & \quad \mbox{} 
+ \frac{15b^3-7Mb^2+17M^2b-M^3}{4(b-M)^{5/2}(b+M)^{11/2}}\, \cos^3\alpha
- \frac{35b^3-29Mb^2+41M^2b-7M^3}{8(b-M)^{5/2}(b+M)^{11/2}}\, \cos^5\alpha \biggr] s^2
+ O(s^3).
\end{align} 
\end{subequations} 

The solution to Eq.~(\ref{sequence_U2}) is of the form
\begin{equation}
U_2 = \frac{u_{-4}}{s^4} + \frac{u_{-3}}{s^3} + \frac{u_{-2}}{s^2} + \frac{u_{-1}}{s}  + u_0 + O(s),
\label{U2dipole} 
\end{equation}
in which the coefficients $u_j$ are functions of $\alpha$ that are required to be regular at $\alpha = 0$ and $\alpha = \pi$. Integrating order-by-order in $s$, we get 
\begin{subequations}
\begin{align}
u_{-4} &= -p^2 \frac{(b-M)^2}{2(b+M)^2}\, \cos^2\alpha, \\
u_{-3} &= -p^2 M \frac{b-M}{(b+M)^3}\, \cos^3\alpha, \\ 
u_{-2} &= -p^2 M \biggl[ \frac{b-M}{2(b+M)^4} - \frac{b-3M}{2(b+M)^4}\, \cos^2\alpha
- \frac{b}{(b+M)^4}\, \cos^4\alpha \biggr], \\
u_{-1} &= p^2 M \biggl[ \frac{M}{(b+M)^5}\, \cos\alpha
- \frac{b-M}{(b+M)^5}\, \cos^2\alpha
+ \frac{1}{2(b-M)(b+M)^3}\, \cos^3\alpha
- \frac{3b^2+M^2}{2(b-M)(b+M)^5}\, \cos^5\alpha \biggr], \\
u_0 &= \mu_0 - p^2 M \biggl[ \frac{1}{2(b+M)^5}\, \cos\alpha
+ \frac{Mb(b-2M)}{(b-M)^2(b+M)^6}\, \cos^2\alpha
- \frac{3b-M}{2(b+M)^6}\, \cos^3\alpha
\nonumber \\ & \quad \mbox{} 
+ \frac{7b^3+4Mb^2+11M^2b+2M^3}{4(b-M)^2(b+M)^6}\, \cos^4\alpha
- \frac{5b(b^2+M^2)}{2(b-M)^2(b+M)^6}\, \cos^6\alpha \biggr],  
\end{align}
\end{subequations}
where $\mu_0$ is an arbitrary constant. The discussion following Eq.~(\ref{U2sol}) in Sec.~\ref{sec:second-order} implies that this solution is complete; there is no need to insert additional terms to account for the distributional source in Eq.~(\ref{sequence_U2}), and there is no need to add particular solutions beyond the constant term $\mu_0$.   

\subsection{String tension} 

With $U$ and $\Phi$ now at hand, the calculation of $\gamma(\rho,z>b)$ proceeds as in Sec.~\ref{subsec:gamma}. We insert the potentials in Eq.~(\ref{dalpha_gamma}), and substitute this within the integral of Eq.~(\ref{gamma_axis}). We obtain
\begin{equation} 
\gamma^{\rm axis} = \epsilon\, \frac{4mM}{(b-M)^{1/2}(b+M)^{3/2}}
- \epsilon^2\, \frac{4p^2 M(2b-M)}{(b+M)^6} + O(\epsilon^3). 
\label{gamma_axis_dipole}  
\end{equation} 
Making the substitution in Eq.~(\ref{tension-intro}), we find that the tension is given by 
\begin{equation} 
T = \epsilon\, \frac{mM}{(b-M)^{1/2}(b+M)^{3/2}} - \epsilon^2\, \frac{2m^2 M^2}{(b-M)(b+M)^3}
- \epsilon^2\, \frac{p^2 M(2b-M)}{(b+M)^6} + O(\epsilon^3).
\label{tension-dipole} 
\end{equation}
As in Eq.~(\ref{tension}), the first term is recognized as $m a$, the particle's mass times its acceleration in the background Schwarzschild spacetime, as given by Eq.~(\ref{acceleration}). The second term is a gravitational self-force correction to $m a$. The last term is (minus) the electromagnetic self-force acting on a point dipole in Schwarzschild spacetime. In the usual Schwarzschild coordinates, with $r_0=b+M$, we have that
\begin{equation}
F_{\rm self}^{\rm dipole} = +\frac{2p^2M}{r_0^5} \biggl( 1 - \frac{3M}{2r_0} \biggr). 
\label{sf-dipole} 
\end{equation} 
As in the case of a point charge, this self-force is repulsive. And as in Sec.~\ref{sec:electric-charge}, all quantities that appear in Eq.~(\ref{tension-dipole}) are well-defined operationally; the coordinate position $b$, in particular, can be related to the regularized redshift by Eq.~(\ref{redshift2}). 

The self-force on an electric dipole in Schwarzschild spacetime was first calculated by L\'eaut\'e and Linet \cite{leaute-linet:84}. Their result differs from ours; they get $(2p^2M/r_0^5) (1 - 5M/2r_0)$.  
L\'eaut\'e and Linet calculate the self-force under the premise that in Eq.~(\ref{Phi-dipole-Schw}), only the last term, $-p \sqrt{f_0} M/(r_0^2 r)$, contributes to the self-force; they dismiss the remaining terms on the grounds that they are singular at the dipole's position.  In the concluding section of their paper, they admit that their treatment is ``simple but not completely rigorous''. The disagreement with our result indicates that the singular terms in $\Phi_1$, when suitably regularized, do have an impact on the self-force. 

\subsection{Magnetic dipole}
\label{sec:magnetic-dipole} 

L\'eaut\'e and Linet \cite{leaute-linet:84} also calculate the self-force on a magnetic dipole in Schwarzschild spacetime, with the same caveat regarding the regularization of singular terms in the electromagnetic field tensor. It is easy to show that this self-force must also be given by Eq.~(\ref{sf-dipole}), with $p$ now interpreted as a magnetic dipole moment. 

As L\'eaut\'e and Linet point out, the field of a magnetic dipole can be obtained from that of an electric dipole by a duality transformation. We therefore begin with the electrostatic potential $\Phi$ of an electric dipole, obtain the corresponding field tensor $F^{\rm el}_{\alpha\beta} = \partial_\alpha A^{\rm el}_\beta - \partial_\beta A^{\rm el}_\alpha$ with $A^{\rm el}_\alpha = -\Phi\, \partial_\alpha t$, and perform the duality transformation 
\begin{equation}
F^{\rm mag}_{\alpha\beta} = -\frac{1}{2} \varepsilon_{\alpha\beta}^{\ \ \ \gamma\delta} F_{\gamma\delta}^{\rm el},
\end{equation}
where $\varepsilon_{\alpha\beta\gamma\delta}$ is the Levi-Civita tensor. We find that both fields give rise to the same distribution of energy-momentum tensor, and that the Einstein field equations for the magnetic dipole take exactly the same form as Eqs.~(\ref{einsteinU}) and (\ref{einsteingamma}). It follows from this observation that the calculations carried out in this section apply unchanged to the case of a magnetic dipole, and we conclude that the self-force is indeed given by Eq.~(\ref{sf-dipole}). This result also is in disagreement with L\'eaut\'e and Linet. 

\section{Scalar charge}
\label{sec:scalar-charge} 

In this section we replace the electric charge of Sec.~\ref{sec:electric-charge} with a scalar charge, and apply the calculational methods of Sec.~\ref{sec:weyl} --- with appropriate modifications --- to this new situation. We shall recover Wiseman's result \cite{wiseman:00}, that the self-force on a static scalar charge in Schwarzschild spacetime vanishes. We shall also obtain additional insights into the problem. 

\subsection{Field equations}

We continue to express the metric as in Eq.~(\ref{weyl-intro}), but we replace the electromagnetic vector potential $A_\alpha$ with a scalar potential $\Psi$. This satisfies
\begin{equation}
\Box \Psi = -4\pi \mu,
\end{equation}
where $\Box := g^{\alpha\beta} \nabla_\alpha \nabla_\beta$ is the covariant wave operator, and
\begin{equation}
\mu = q \int \frac{\delta\bigl( x - X(\tau)\bigr)}{\sqrt{-g}}\, d\tau
\end{equation}
is the scalar charge density, with $q$ denoting the charge of a particle moving on a world line described by $X^\alpha(\tau)$; $\tau$ is proper time on the world line. In the case of static particle at $\rho = 0$ and $z = b$, we have that
\begin{equation}
\mu = q e^{-(3U+2\gamma)}\, \delta(\bm{x}-\bm{b}),
\end{equation}
where $\delta(\bm{x}-\bm{b}) := \rho^{-1} \delta(\rho) \delta(z-b) \delta(\phi)$ is a three-dimensional delta function.

The Einstein field equations are
\begin{equation}
G_{\alpha\beta} - 2\partial_\alpha \Psi \partial_\beta \Psi
+ g_{\alpha\beta} g^{\mu\nu} \partial_\mu \Psi \partial_\nu \Psi = 8\pi T_{\alpha\beta},
\end{equation}
where $G_{\alpha\beta}$ is the Einstein tensor, the remaining terms on the left-hand side make up the energy-momentum tensor of the scalar field, and $T_{\alpha\beta}$ is the particle's energy-momentum tensor, as given by Eqs.~(\ref{Tpart1}) and (\ref{Tpart2}).

The explicit form of the field equations is
\begin{subequations}
\label{scalar-einstein1} 
\begin{align} 
\nabla^2 \Psi &= -4\pi q e^{-U}\, \delta(\bm{x}-\bm{b}), \\
\nabla^2 U &= -4\pi m e^{-U}\, \delta(\bm{x}-\bm{b}), 
\end{align}
\end{subequations}
where $\nabla^2 := \partial_{\rho\rho} + \rho^{-1} \partial_\rho + \partial_{zz}$, and
\begin{subequations}
\label{scalar-einstein2} 
\begin{align}
\frac{1}{\rho} \partial_\rho \gamma &= (\partial_\rho U)^2 - (\partial_z U)^2
+ (\partial_\rho \Psi)^2 - (\partial_z \Psi)^2, \\
\frac{1}{\rho} \partial_z \gamma &= 2 (\partial_\rho U) (\partial_z U)
+ 2 (\partial_\rho \Phi) (\partial_z \Phi).
\end{align}
\end{subequations}
It is remarkable that apart from the factor of $e^{-U}$ that comes with the delta functions, the field equations for $\Psi$ and $U$ are completely independent, and that they each take the form of a simple Poisson equation.

\subsection{Solution} 

We might, in the spirit of Sec.~\ref{sec:electric-charge}, integrate Eqs.~(\ref{scalar-einstein1}) by inserting the perturbative expansions $U = U_0 + \epsilon U_1 + \epsilon^2 U_2 + O(\epsilon^3)$ and $\Psi = \epsilon \Psi_1 + \epsilon^2 \Psi_2 + O(\epsilon^3)$. In this approach, $U_0$ would be the Schwarzschild potential of Eq.~(\ref{sch_metric}), and at first order in $\epsilon$, the factor of $e^{-U}$ multiplying $\delta(\bm{x}-\bm{b})$ would be replaced by $(b-M)^{1/2}/(b+M)^{1/2}$, the value of $e^{-U_0}$ at $\rho=0$, $z=b$. The solutions to Eqs.~(\ref{scalar-einstein1}) would then be
\begin{equation}
\Psi_1 = q \sqrt{\frac{b-M}{b+M}} \frac{1}{D}, \qquad
U_1 = m \sqrt{\frac{b-M}{b+M}} \frac{1}{D},
\end{equation}
where $D := \sqrt{\rho^2 + (z-b)^2}$. At second order in $\epsilon$ we would regularize the distributional source terms as in Sec.~\ref{sec:second-order}, and replace the factor multiplying $\delta(\bm{x}-\bm{b})$ by its {\it Hadamard partie finie}, which vanishes. We would thereby obtain $\Psi_2 = 0 = U_2$, and conclude that the perturbative expansion terminates at order $\epsilon$.

The purpose of this rather belabored discussion is to bring home the point that 
\begin{equation}
\Psi = \Psi_1, \qquad 
U = U_0 + U_1
\end{equation}
can be viewed as an {\it exact solution} to the Einstein-scalar equations. Viewed perturbatively, the solution describes a point particle of mass $m$ and scalar charge $q$ held in place outside a Schwarzschild black hole of mass $M$. Viewed as an exact solution, the point particle is replaced by a Curzon-Chazy object of the same mass and charge. In this interpretation, the distributional sources can again be made meaningful with the help of Hadamard's prescription. A better option, however, is simply to eliminate the source terms all together, take $\Psi_1$ and $U_1$ to satisfy Laplace's equation, and adopt for them a specific singular behavior at $\rho = 0$, $z=b$. In this view, the Laplace equations are meant to apply everywhere, except at the singularity. 

Equations (\ref{scalar-einstein2}) can be integrated exactly for our potentials $\Psi$ and $U$. We find that 
\begin{equation}
\gamma = \gamma_0 + \gamma_1 + \gamma_2,
\end{equation}
where $\gamma_0$ is the Schwarzschild expression of Eq.~(\ref{sch_metric}), $\gamma_1$ is given by Eq.~(\ref{gamma1}), and
\begin{equation}
\gamma_2 = -\frac{1}{2} \Bigl( m^2 + q^2 \Bigr) \frac{b-M}{b+M} \frac{\rho^2}{D^4}.
\end{equation}
Remarkably, while $\gamma_1$ is nonzero when $\rho = 0$ and $z > b$ (or $z < -M$), $\gamma_2$ vanishes everywhere on the $z$-axis (except at the singular point).

\subsection{String tension} 

On the axis, for $z > b$ or $z < -M$, we find that
\begin{equation}
\gamma^{\rm axis} =\frac{4mM}{(b-M)^{1/2}(b+M)^{3/2}},
\end{equation}
and it follows that the string's tension is
\begin{equation}
T = \frac{1}{4} \Biggl\{ 1 - \exp\biggl[-\frac{4mM}{(b-M)^{1/2}(b+M)^{3/2}} \biggr] \Biggr\}.
\label{tension-scalar}
\end{equation}
The fact that $T$ is independent of $q$ is the statement that there is no scalar self-force contribution to the force required to hold the particle in place. It was first demonstrated by Wiseman \cite{wiseman:00} that the scalar self-force vanishes in the Schwarzschild spacetime; we have here a nonperturbative extension of his  result.  

\subsection{Schwarzschild coordinates} 

In the usual Schwarzschild coordinates, the metric of our scalarized Curzon-Chazy object is 
\begin{equation}
ds^2 = -e^{-2U_1} f\, dt^2 + e^{2(U_1 + \gamma_1 + \gamma_2)} \bigl( f^{-1}\, dr^2 + r^2\, d\theta^2 \bigr)
+ e^{2U_1} r^2\sin^2\theta\, d\phi^2,
\end{equation}
where $f := 1-2M/r$, $U_1 = m\sqrt{f_0}/D$, $\gamma_1$ is given by Eq.~(\ref{g1_Schw}),
\begin{equation}
\gamma_2 = -\frac{1}{2} (m^2+q^2) f_0 \frac{r^2 f \sin^2\theta}{D^4},
\end{equation}
and where $D$ now takes the form of Eq.~(\ref{D_Schw}). The object is situated at $r = r_0 = b+M$, $\theta=0$, and we have that $f_0 := 1-2M/r_0$. The scalar potential is $\Psi = q\sqrt{f_0}/D$, which is recognized as Wiseman's solution \cite{wiseman:00} for a point scalar charge in the Schwarzschild spacetime.

\subsection{Properties of the deformed black hole} 

The event horizon of the perturbed black hole is still situated at $r = 2M$, and because $\gamma_2 = 0$ on the horizon, its intrinsic geometry is described by the induced metric
\begin{equation}
ds^2 = 4M^2 e^{2U_1} \bigl( e^{2\gamma_1}\, d\theta^2 + \sin^2\theta\, d\phi^2 \bigr),
\end{equation}
with $U_1(r=2M)$ and $\gamma_1(r=2M)$ still given by Eq.~(\ref{horizon_values}). The horizon area is 
\begin{equation}
A = 16\pi M^2 \exp\biggl( \frac{2m}{r_0\sqrt{f_0}} \biggr),  
\end{equation} 
and the surface gravity is calculated to be 
\begin{equation}
\kappa = \frac{1}{4M} \exp \biggl(-\frac{2m}{r_0\sqrt{f_0}} \biggr). 
\end{equation}
It follows from this that $M_{\rm Smarr} := \kappa A/(4\pi) = M$. The first law of black-hole mechanics continues to take the form of $d M_{\rm tot} = (\kappa/8\pi)\, dA - \lambda\, dT$, where $M_{\rm tot} := M + m\sqrt{f_0}$ is the total mass, $T$ is the tension of Eq.~(\ref{tension-scalar}), and
\begin{equation}
\lambda := r_0 \exp\biggl( \frac{4 m M}{r_0^2 \sqrt{f_0}} \biggr)
\end{equation}
is the string's thermodynamic length.  

\subsection{Regularized redshift}

We can, as in Sec.~\ref{sec:first-order}, define a regularized redshift by removing the singular factor $e^{-U_1}$ from the accounting of the gravitational potential; in this prescription, the local gravity of the scalarized Curzon-Chazy object is simply taken out of the redshift formula. The prescription returns
\begin{equation}
\gothf_{\rm reg} = \sqrt{f_0}, \qquad
r_0/M = \frac{2}{1 - \gothf_{\rm reg}^2},
\label{redshift_scalar} 
\end{equation}
in agreement with Eqs.~(\ref{redshift1}) and (\ref{redshift2}), respectively. There is a difference, however: the relation between $\gothf_{\rm reg}$ and $r_0$ is now meant to be exact, instead of an approximation through order $\epsilon$. 

Because $e^{-U_1}$ contains terms of all orders in $s^{-1}$ (with $s$ denoting the coordinate distance from the Curzon-Chazy object), the prescription of Eq.~(\ref{redshift_scalar}) can no longer be related to a regularization procedure in which $e^{-U_1} f^{1/2}$ is replaced by its {\it Hadamard partie finie}. The relationship, however, is recovered when one retreats to a perturbative interpretation of our results, and take them to apply only through first order in an expansion in powers of $\epsilon$.  

\subsection{Scalar dipole} 

The developments of this section can be generalized to other configurations for the scalar field. For example, the Curzon-Chazy object could be endowed with a scalar dipole moment instead of a scalar charge, and the corresponding dipole solution to Laplace's equation could be adopted for $\Psi$. We have gone through this exercise, and find that the scalar self-force vanishes in this case also; the string's tension continues to be given by Eq.~(\ref{tension-scalar}). We omit the details of this calculation here, because the resulting expression for $\gamma_2$ is lengthy and not terribly illuminating. The only important point is that in this case also, $\gamma_2$ vanishes on the axis, so that the string's tension comes entirely from $\gamma_1$. The tension, therefore, depends on $m$ but is independent of the scalar dipole moment.    

\section{Force on extended objects} 
\label{sec:extended} 

In the preceding sections of the paper we calculated the force required to hold a pointlike object in place, as measured by the tension in a Nambu-Goto string attached to the object. Here we generalize the method to handle extended objects, following a technique devised by Weinstein \cite{weinstein:90} to calculate the stress in a strut that holds two Schwarzschild black holes apart. To introduce the method we revisit the case of two black holes, replacing the strut by strings, and then examine the case of two Janis-Newman-Winicour objects \cite{janis-newman-winicour:68}, naked singularities which are individually described by scalarized Schwarzschild solutions.  

\subsection{Schwarzschild black holes}
\label{sec:twohole} 

Following Weinstein, we consider two Schwarzschild black holes held apart by a pair of Nambu-Goto strings; each black hole is attached to a string, which extends from the black hole to infinity. The first black hole has a mass $M$, it is situated at $\rho = 0$, $z = 0$, and it comes with a gravitational potential $U_0$ given by 
\begin{equation} 
e^{-2U_0} = \frac{R_++R_--2M}{R_++R_-+2M}, \qquad 
R_\pm := \sqrt{\rho^2 + (z\pm M)^2}. 
\end{equation} 
The second black hole has a mass $m$, it is situated at $\rho=0$, $z=b$, and its gravitational potential $U_1$ is
\begin{equation} 
e^{-2U_1} = \frac{r_++r_--2m}{r_++r_-+2m}, \qquad 
r_\pm := \sqrt{\rho^2 + (z-b \pm m)^2}. 
\end{equation} 
We assume that $b > M + m$. The potentials $U_0$ and $U_1$ are each solutions to Laplace's equation in cylindrical coordinates, and we construct the two-hole spacetime by superposing these solutions: $U = U_0 + U_1$. It should be kept in mind that while the interpretation of $M$ and $m$ as mass parameters was sound in the context of the individual solutions, it does not hold up in the context of the superposition. In particular, according to $U = U_0 + U_1$, the total mass in the spacetime is simply $M + m$, and this must account for the system's gravitational binding energy. This implies that $M$ actually represents the mass of the first black hole minus a fraction of the binding energy, and that $m$ incorporates the remaining fraction of this energy.   

To obtain $\gamma$ we write 
\begin{equation} 
\gamma = \gamma_{00} + \gamma_{01} + \gamma_{11}, 
\end{equation} 
and invoke Eqs.~(\ref{einsteingamma}) to obtain 
\begin{subequations} 
\begin{align} 
\frac{1}{\rho} \partial_\rho \gamma_{00} &=  (\partial_\rho U_0)^2 - (\partial_z U_0)^2, \\ 
\frac{1}{\rho} \partial_\rho \gamma_{01} &= 2(\partial_\rho U_0) (\partial_\rho U_1) 
- 2(\partial_z U_0) (\partial_z U_1), 
\label{g01_rho} \\ 
\frac{1}{\rho} \partial_\rho \gamma_{11} &= (\partial_\rho U_1)^2 - (\partial_z U_1)^2 
\end{align} 
\end{subequations} 
and 
\begin{subequations} 
\begin{align} 
\frac{1}{\rho} \partial_z \gamma_{00} &= 2(\partial_\rho U_0) (\partial_z U_0), \\  
\frac{1}{\rho} \partial_z \gamma_{01} &= 2(\partial_\rho U_0) (\partial_z U_1) 
+ 2 (\partial_z U_0)(\partial_\rho U_1), 
\label{g01_z} \\ 
\frac{1}{\rho} \partial_z \gamma_{11} &= 2(\partial_\rho U_1) (\partial_z U_1).   
\end{align} 
\end{subequations} 
The notation should be clear: $\gamma_{00}$, as given by Eq.~(\ref{sch_metric}), is the potential associated with the black hole of mass $M$ as if it were isolated, and $\gamma_{11}$ is similarly associated with the black hole of mass $m$; $\gamma_{01}$ results from the interaction between black holes. We have that  $\gamma_{00}$ and $\gamma_{11}$ both vanish on the $z$-axis (except in the segments occupied by the black holes), and therefore do not contribute to the angular deficit nor the string's tension. To calculate the tension, we may focus our attention entirely on $\gamma_{01}$. 

We take $\gamma$ to vanish on the axis segment between the black holes (for $M < z < b-m$), and we calculate its value above the black hole of mass $m$ (for $z > b+m$) with the help of the line integral 
\begin{equation} 
\gamma^{\rm axis} = \int_C (\partial_\rho \gamma_{01}\, d\rho 
+ \partial_z \gamma_{01}\, dz), 
\end{equation} 
where $C$ is any path away from the axis that links a point $P$ below $z=b-m$ to a point $Q$ above $z=b+m$. Equations (\ref{g01_rho}) and (\ref{g01_z}) and straightforward manipulations allow us to express this as 
\begin{equation} 
\gamma^{\rm axis} = 2\int_C \rho \partial_\rho U_0\, dU_1 
+ 2\int_C (\partial_z U_0) \bigl[ (1 + \rho \partial_\rho U_1)\, dz - \rho \partial_z U_1\, d\rho \bigr] 
+ 2\int_C \partial_\rho U_0\, d\rho 
-2 \bigl[ U_0(Q) - U_0(P) \bigr]. 
\end{equation} 
For $C$ we choose the elliptical path described by $\rho = s \sin\alpha$ and $z = b - (m^2+s^2)^{1/2} \cos\alpha$, where $s$ is constant and $0 \leq \alpha \leq \pi$; because $\gamma^{\rm axis}$ is independent of $C$, and therefore of $s$, it is sufficient to calculate each integral in the limit $s \to 0$. We observe that $r_\pm = (m^2+s^2)^{1/2} \mp m\cos\alpha$ on $C$, and it follows that $U_1$ is constant on the adopted path; the first integral vanishes. In the second integral, we find that the quantity between square brackets evaluates to $[(m^2+s^2)^{1/2}-m]\sin\alpha\, d\alpha$, and that it scales as $O(s^2)$ in the limit $s \to 0$; the second integral makes no contribution in the limit. The same conclusion applies to the third integral, which scales as $O(s)$. The final result is that 
\begin{equation} 
\gamma^{\rm axis} = -2 \bigl[ U_0(\rho=0,z=b+m) - U_0(\rho=0,z=b-m) \bigr], 
\end{equation} 
and making the substitutions in $U_0(\rho,z)$, we obtain
\begin{equation} 
\gamma^{\rm axis} = \ln \frac{b^2 - (M-m)^2}{b^2 - (M+m)^2}. 
\end{equation} 

The string's tension is obtained by inserting our result within Eq.~(\ref{tension-intro}). This gives 
\begin{equation} 
T = \frac{mM}{b^2 - (M-m)^2}. 
\label{tension-twohole} 
\end{equation} 
The value reported by Weinstein \cite{weinstein:90} has $(M+m)^2$ in the denominator. The discrepancy is real, but Weinstein's result is nevertheless correct. Instead of a string attached to each black hole, Weinstein puts a strut between the black holes, and calculates the stress in the strut. With $\gamma(\rho,z)$ assumed to be zero for $z > b+m$, the value for $M < z < b-m$ is $-\gamma^{\rm axis}$, and the stress is equal to $\frac{1}{4}[ \exp(\gamma^{\rm axis}) - 1 ] = mM/[b^2 - (M+m)^2]$, in agreement with Weinstein's result.      

In the case of the pointlike objects of Secs.~\ref{sec:electric-charge}, \ref{sec:electric-dipole}, and \ref{sec:scalar-charge} we were able to give operational meanings to the quantities that appear in the string's tension. For example, we recall that $m$ was related to the particle's energy-momentum tensor, that $M$ was identified with the Smarr mass of the black hole, and that $b$ was related to a regularized redshift. It should also be possible to assign such operational meanings to $m$, $M$, and $b$ in the context of the two-hole spacetime. For example, it is plausible that $M$ and $m$ could be expressed in terms of geometric quantities defined on the horizon of each black hole, and that $b$ could be related to the proper spatial distance between horizons. These identifications, however, would require a complete knowledge of the metric. Because $\gamma$ is known only on the $z$-axis, our knowledge is incomplete, and the operational meaning of $m$, $M$, and $b$ remains unknown. In view of this, the interpretation of Eq.~(\ref{tension-twohole}) must remain ambiguous.    

\subsection{Janis-Newman-Winicour objects} 
 
A Janis-Newman-Winicour (JNW) object \cite{janis-newman-winicour:68} is a naked singularity with mass $M$ and scalar charge $Q$. It is described by an exact solution to the Einstein-scalar equations of Sec.~\ref{sec:scalar-charge}, in which the distributional source terms are set to zero. The solution is a scalarized version of the Schwarzschild metric, obtained by setting 
\begin{equation} 
U = \frac{1}{\Lambda} \chi, \qquad 
\Psi = \frac{Q}{\Lambda M} \chi, \qquad 
\Lambda = \sqrt{1 + Q^2/M^2}, 
\label{JNWtransf} 
\end{equation} 
where $\chi(\rho,z)$ is an auxiliary potential that satisfies $\nabla^2 \chi = 0$. The remaining field equations (\ref{scalar-einstein2}) become 
\begin{subequations}
\begin{align}
\frac{1}{\rho} \partial_\rho \gamma &= (\partial_\rho \chi)^2 - (\partial_z \chi)^2, \\
\frac{1}{\rho} \partial_z \gamma &= 2 (\partial_\rho \chi) (\partial_z \chi).
\end{align}
\end{subequations}
The equations for $\chi$ and $\gamma$ are formally identical to the Einstein field equations in vacuum, and the transformation of Eq.~(\ref{JNWtransf}) can therefore be exploited to turn a pure-gravity solution into a solution of the Einstein-scalar equations. 

The JNW metric and scalar potential are given by 
\begin{equation} 
e^{-2\chi} = \frac{R_+ + R_- - 2\Lambda M}{R_+ + R_- + 2\Lambda M}, \qquad 
e^{2\gamma} = \frac{(R_+ + R_-)^2 - 4(\Lambda M)^2}{4 R_+ R_-}, 
\end{equation} 
where $R_\pm := [\rho^2 + (z\pm \Lambda M)^2]^{1/2}$. In these expressions, the mass parameter $M$ is multiplied by $\Lambda$ to account for the compensating factor in Eq.~(\ref{JNWtransf}). The transformation 
\begin{equation} 
\rho = \sqrt{ (R+M)^2 - (\Lambda M)^2 }\sin\theta, \qquad 
z = (R + M) \cos\theta
\end{equation} 
to new coordinates $(R,\theta)$ brings the metric and scalar potential to the forms originally given by Janis, Newman, and Winicour. It is useful to note that in the new coordinates, $R_\pm = R + M \pm \Lambda M \cos\theta$ and $d\rho^2 + dz^2 = R_+ R_-\{ [(R+M)^2 - (\Lambda M)^2]^{-1}\, dR^2 + d\theta^2 \}$. 

We now construct a spacetime that contains two such JNW objects. The first has a mass $M$ and scalar charge $Q$, and is situated at $\rho = 0$, $z=0$; its potentials $U_0$ and $\Psi_0$ are those given previously. The second has a mass $m$ and charge $q$, and is situated at $\rho = 0$, $z = b$. Its potentials are 
\begin{equation} 
U_1 = \frac{1}{\lambda} \chi_1, \qquad 
\Psi_1 = \frac{q}{\lambda m} \chi_1, \qquad 
\lambda = \sqrt{1 + q^2/m^2},   
\end{equation} 
with 
\begin{equation} 
e^{-2\chi_1} = \frac{r_+ + r_- - 2\lambda m}{r_+ + r_- + 2\lambda m} 
\end{equation} 
and $r_\pm := \sqrt{\rho^2 + (z-b \pm \lambda m)^2}$. We assume that $b > \Lambda M + \lambda m$. We write $U = U_0 + U_1$, $\Psi = \Psi_0 + \Psi_1$, and $\gamma = \gamma_{00} + \gamma_{01} + \gamma_{11}$. As we saw previously in the case of the superposed Schwarzschild solutions, $\gamma_{00}$ and $\gamma_{11}$ correspond to the objects in isolation, while $\gamma_{01}$ accounts for their interaction. Its governing equations are 
\begin{subequations} 
\begin{align} 
\frac{1}{\rho} \partial_\rho \gamma_{01} &= \frac{2}{\lambda \Lambda} \biggl( 1 + \frac{qQ}{mM} \biggr)
\Bigl[ (\partial_\rho \chi_0) (\partial_\rho \chi_1) - (\partial_z \chi_0) (\partial_z \chi_1) \Bigr], \\  
\frac{1}{\rho} \partial_\rho \gamma_{01} &= \frac{2}{\lambda \Lambda} \biggl( 1 + \frac{qQ}{mM} \biggr)
\Bigl[ (\partial_\rho \chi_0) (\partial_z \chi_1) + (\partial_z \chi_0) (\partial_\rho \chi_1) \Bigr]. 
\end{align} 
\end{subequations} 
Repeating the manipulations of the preceding subsection, we calculate $\gamma^{\rm axis}$ with the help of a line integral, with a path $C$ now described by $\rho = s \sin\alpha$ and $z = b - [(\lambda m)^2 + s^2]^{1/2} \cos\alpha$, with $s = \mbox{constant}$ and $0 \leq \alpha \leq \pi$. We get
\begin{equation} 
\gamma^{\rm axis} = -\frac{2}{\lambda \Lambda} \biggl( 1 + \frac{qQ}{mM} \biggr)
\Bigl[ \chi_0(\rho=0,z=b+\lambda m) - \chi_0(\rho=0,z=b-\lambda m) \Bigr], 
\end{equation} 
or 
\begin{equation} 
\gamma^{\rm axis} = \frac{1}{\lambda \Lambda} \biggl( 1 + \frac{qQ}{mM} \biggr)
\ln \frac{ b^2 - (\Lambda M - \lambda m)^2 }{ b^2 - (\Lambda M + \lambda m)^2 }. 
\end{equation} 
The string's tension is then obtained by inserting this within Eq.~(\ref{tension-intro}). These results come with the same warnings as in Sec.~\ref{sec:twohole}: a precise operational interpretation of the parameters $M$, $m$, and $b$ must await a complete determination of the metric. The scalar charges $Q$ and $q$, however, can be defined operationally in terms of the scalar potential.  

It is noteworthy that $\gamma^{\rm axis}$ and $T$ both vanish when $mM = -qQ$, that is, when the gravitational attraction of the JNW objects is balanced by their scalar repulsion --- unlike charges repel in this theory. When $b \gg \Lambda M + \lambda m$, the expressions simplify to 
\begin{equation} 
\gamma^{\rm axis} = \frac{4(mM+qQ)}{b^2} \biggl[ 
1 + \frac{m^2+q^2+M^2+Q^2}{b^2} + O(b^{-4}) \biggr] 
\end{equation} 
and 
\begin{equation} 
T = \frac{mM+qQ}{b^2} \biggl[ 
1 + \frac{m^2+q^2-2(mM+qQ)+M^2+Q^2}{b^2} + O(b^{-4}) \biggr]. 
\end{equation} 

\begin{acknowledgments} 
This work was supported by the Natural Sciences and Engineering
Research Council of Canada.  
\end{acknowledgments} 

\appendix 
\section{Black hole and particle views} 
\label{sec:BHPartView} 

In Sec.~\ref{sec:electric-charge} we examined a point particle of mass $m$ and electric charge $q$ held in place outside a Schwarzschild black hole of mass $M$, and calculated the metric and vector potential created by this particle-black hole system. The metric and vector potential were presented as expansions in powers of $\epsilon$, a book-keeping parameter that keeps track of the powers of $m$ and $q$. We saw that while the order-$\epsilon$ potentials could be obtained globally and expressed in closed forms, only local expansions could be given for the order-$\epsilon^2$ potentials. Moreover, we saw that the local solutions for the order-$\epsilon^2$ potentials were not unique, and involved arbitrary constants.    

In this Appendix we aim to obtain additional information about the local solutions, and we achieve this by exploiting the method of matched asymptotic expansions. We construct two versions of $U$ and $\Phi$, one reflecting the black-hole view of Sec.~\ref{sec:electric-charge}, the other reflecting a particle view to be developed here. In the black-hole view, we have a Schwarzschild black hole perturbed by a point particle, and the expansion parameter is $\epsilon$. In the particle view, we have a charged particle --- modelled as a Reissner-Nordstr\"om (RN) field --- perturbed by the tidal environment provided by the black hole, and the expansion parameter is $M/b^{\ell+1}$, which measures the strength of the tidal field for a multipole of order $\ell$. Each version of $U$ and $\Phi$ comes with unknown constants, and these are approximately determined by demanding that both versions be mutually compatible when $\epsilon$ and $M/b^{\ell+1}$ are simultaneously small.

\subsection{Tidal perturbation of a RN field}
\label{sec:PartView}

In the particle view, the charged particle is modelled as a tidally deformed Reissner-Nordstr\"om (RN) field; it could be a black hole (when $m > |q|$) or a naked singularity (when $m < |q|$). The source of the tidal deformation is a black hole of mass $M$, situated at a distance $b$ from the particle.  

We express the metric of a tidally deformed RN field as 
\begin{equation} 
ds^2 = -e^{-2\delta U} f\, d\bar{t}^2 + e^{2(\delta U + \delta \gamma)} (f^{-1}\, dr^2 + r^2\, d\theta^2) 
+ e^{2\delta U} r^2\sin^2\theta\, d\phi^2, 
\end{equation} 
where $f := 1 - 2m/r + q^2/r^2$, and where $\delta U(r,\theta)$, $\delta \gamma(r,\theta)$ are the gravitational perturbations. The vector potential is expressed as 
\begin{equation} 
A_\alpha = -(q/r + \delta \Phi) \partial_\alpha \bar{t}, 
\end{equation} 
where $\delta \Phi(r,\theta)$ is the electromagnetic perturbation. It is assumed that the tidal perturbation is static and axisymmetric. We place an overbar on the time coordinate to distinguish it from the $t$ used in the main text; we shall see that they are related by a scaling factor. 

The field equations are decoupled by writing
\begin{equation} 
\delta \Phi = (m - q^2/r) (\sqrt{f} A) + q f B, \qquad 
\delta U = (m-q^2/r) B + q (\sqrt{f} A), 
\label{decoupling} 
\end{equation}
and the auxiliary functions $A$ and $B$ are decomposed as 
\begin{equation} 
A(r,\theta) = \sum_{\ell=0}^\infty A_\ell(r) P_\ell(\cos\theta), \qquad 
B(r,\theta) = \sum_{\ell=0}^\infty B_\ell(r) P_\ell(\cos\theta), 
\label{AB_series} 
\end{equation} 
where $P_\ell(\cos\theta)$ are Legendre polynomials. The field equations become 
\begin{subequations}
\begin{align} 
0 &= r^2 f \frac{d^2 A_\ell}{dr^2} + 2(r-m) \frac{dA_\ell}{dr} - \biggl[ \ell(\ell+1)
+ \frac{m^2-q^2}{r^2 f} \biggr] A_\ell, \\
0 &= r^2 f \frac{d^2 B_\ell}{dr^2} + 2(r-m) \frac{dB_\ell}{dr} - \ell(\ell+1) B_\ell. 
\end{align}
\end{subequations}
To represent a tidal perturbation we adopt the {\it growing solutions} to these equations, those that behave as $r^\ell$ when $r$ is large. We exclude decaying solutions, those that behave as $r^{-(\ell+1)}$ when $r$ is large, for the following reasons: When the RN field describes a black hole (when $m > |q|$), the decaying solutions diverge at the horizon, and they are rejected on the grounds that the perturbation should be well-behaved there. When the RN field describes a naked singularity (when $m < |q|$), the decaying solutions would be associated with intrinsic multipole moments, and these are set to zero to keep the particle as spherical as possible.

For $\ell \neq 0$, the growing solutions to the perturbation equations are  
\begin{equation} 
A_\ell = a_\ell P^1_\ell\bigl( (r-m)/\lambda \bigr), \qquad  
B_\ell = b_\ell P_\ell\bigl( (r-m)/\lambda \bigr), 
\end{equation} 
where $P_\ell$ are Legendre polynomials, $P_\ell^m$ are associated Legendre functions, $a_\ell$, $b_\ell$ are arbitrary constants, and $\lambda := \sqrt{m^2-q^2}$ --- this parameter can be real or imaginary. For small values of $\ell$ we have the explicit expressions 
\begin{subequations}
\label{tidal_multipoles} 
\begin{align} 
\delta U_1 &= \biggl[ u_1 \biggl( r - m + \frac{q^2}{r} \biggr) - p_1 q \biggr] P_1(\cos\theta), \\
\delta \Phi_1 &= f ( p_1 r - u_1 q ) P_1(\cos\theta), \\
\delta U_2 &= \biggl[ u_2 \biggl( r^2 - 2m r + \frac{2}{3} m^2 + q^2 - \frac{2}{3} \frac{m q^2}{r} \biggr)
          - p_2 q \biggl( r - \frac{4}{3}m + \frac{1}{3} \frac{q^2}{r} \biggr) \biggr] P_2(\cos\theta), \\ 
\delta \Phi_2 &= f \biggl[ p_2 \biggl( r^2 - m r + \frac{1}{3} q^2 \biggr)
          - u_2 q \biggl( r - \frac{2}{3}m \biggr) \biggr] P_2(\cos\theta). 
\end{align}
\end{subequations}
The constants $b_\ell$ and $a_\ell$ were eliminated in favor of $u_\ell$ and $p_\ell$; these are defined so that $u_\ell$ is the coefficient in front of $r^\ell P_\ell(\cos\theta)$ in $\delta U_\ell$, while $p_\ell$ is the coefficient in front of $r^\ell P_\ell(\cos\theta)$ in $\delta \Phi_\ell$.

For $\ell = 0$ the general solution to the perturbation equations is characterized by three constants of integration, which we denote $u_0$, $p_0$, and $a_0$; a fourth constant is set to zero to ensure that the solution is regular.\footnote{The discarded solution comes with $B_0 = \ln[(r-m+\lambda)/(r-m-\lambda)]$. When the RN field describes a black hole (when $m > |q|$), $B_0$ is singular on the horizon, at $r=m+\lambda$. When the field describes a naked singularity (when $m < |q|$), $B_0$ is purely imaginary, and it is multiplied by an imaginary constant to return a real solution; but this solution possesses a discontinuous derivative at $r = m$.}  The solution is
\begin{subequations}
\begin{align}
\delta U_0 &= u_0 + a_0 \frac{q}{r}, \\
\delta \Phi_0 &= p_0 - u_0 \frac{q}{r} + a_0 \biggl( \frac{m}{r} - \frac{q^2}{r^2} \biggr).
\end{align}
\end{subequations} 
The constant $p_0$ changes $\Phi$ by an irrelevant constant, and $p_0$ can be set equal to zero without loss of generality. The constants $a_0$ and $u_0$ can be related to shifts in the particle's mass and charge parameters. We set these to zero, because the tidal deformation cannot change the particle's charge, and can only change its  mass at second order in the perturbation. We conclude that the $\ell = 0$ contribution to $\delta U$ and $\delta \Phi$ must be eliminated.

The complete perturbation is
\begin{equation}
\delta U = \sum_{\ell=1}^\infty \delta U_\ell, \qquad
\delta \Phi = \sum_{\ell=1}^\infty \delta \Phi_\ell.
\end{equation}
While the sums implicate an infinite number of terms, we shall see below that a truncation through $\ell = 2$ will be sufficient for our purposes. 

\subsection{Transformation to local coordinates} 

To compare the potentials of the particle view to those of the black-hole view, we must transform the coordinates to a local system $(\bar{s},\alpha)$ that will be simply related to the local polar coordinates employed previously. 
We note first that the transformation to Weyl coordinates is $\bar{\rho} = r\sqrt{f}\sin\theta$ and $\bar{z} = (r-m)\cos\theta$. The additional transformation to local coordinates is then $\bar{\rho} = \bar{s}\sin\alpha$ and $\bar{z} = \bar{s}\cos\alpha$. We introduce 
\begin{equation} 
D_\pm := \sqrt{\bar{\rho}^2 + (\bar{z}\pm \lambda)^2} = \sqrt{\bar{s}^2 \pm 2\lambda\bar{s}\cos\alpha + \lambda^2}, 
\end{equation} 
where, we recall, $\lambda := \sqrt{m^2-q^2}$. The inverse transformation is 
\begin{equation} 
r = \frac{1}{2} (D_+ + D_- + 2m), \qquad 
\cos\theta = \frac{2\bar{z}}{D_+ + D_-} = \frac{2\bar{s}\cos\alpha}{D_+ + D_-}. 
\end{equation} 
Expanding in powers of $\lambda^2/\bar{s}^2$, this is 
\begin{subequations} 
\label{inverse_local} 
\begin{align} 
r &= \bar{s} \biggl[ 1 + \frac{m}{\bar{s}} + \frac{\lambda^2 \sin^2\alpha}{2\bar{s}^2} 
+ \frac{\lambda^4 (5\cos^2\alpha-1) \sin^2\alpha}{8\bar{s}^4} + O(\lambda^6/\bar{s}^6) \biggr], \\ 
\theta &= \alpha + \frac{\lambda^2\sin\alpha\cos\alpha}{2\bar{s}^2} 
+ \frac{3\lambda^4 (2\cos^2\alpha - 1) \sin\alpha\cos\alpha}{8 \bar{s}^4} + O(\lambda^6/\bar{s}^6). 
\end{align} 
\end{subequations} 
If we introduce 
\begin{equation} 
e^{-2\bar{U}_0} := f = \frac{ (D_++D_-)^2 - 4\lambda^2 }{(D_++D_-+2m)^2}, \qquad 
e^{2\bar{\gamma}_0} := \frac{ (D_++D_-)^2 - 4\lambda^2 }{4D_+ D_-}, 
\end{equation} 
we find that $f^{-1}\, dr^2 + r^2\, d\theta^2 = e^{2(\bar{U}_0+\bar{\gamma}_0)}(d\bar{\rho}^2 + d\bar{z}^2) = e^{2(\bar{U}_0+\bar{\gamma}_0)}( d\bar{s}^2 + \bar{s}^2\, d\alpha^2)$. 

The metric becomes 
\begin{equation} 
ds^2 = -e^{-2\bar{U}}\, d\bar{t}^2 + e^{2(\bar{U}+\bar{\gamma})} (d\bar{s}^2 + \bar{s}^2\, d\alpha^2) + e^{2\bar{U}} \bar{s}^2\sin^2\alpha\, d\phi^2 
\label{metricPart} 
\end{equation} 
in the local coordinates, where $\bar{U} := \bar{U}_0 + \delta U$ and $\bar{\gamma} = \bar{\gamma}_0 + \delta\gamma$. The potentials $\bar{U}_0$ and $\bar{\gamma}_0$ are recognized as the Weyl expression of the RN solution. The vector potential is
\begin{equation} 
A_\alpha = -\bar{\Phi}\, \partial_\alpha \bar{t},
\end{equation} 
where $\bar{\Phi} = \bar{\Phi}_0 + \delta \Phi$, with 
\begin{equation} 
\bar{\Phi}_0 := \frac{2q}{D_+ + D_- + 2m} 
\end{equation} 
denoting the RN electrostatic potential. 

\subsection{Asymptotic matching} 

The metric and vector potential constructed in Sec.~\ref{sec:electric-charge}, and the ones obtained in this Appendix, give distinct approximate representations of the same physical situation, and they must be mutually compatible. We recall that in the black-hole view, we have a Schwarzschild black hole perturbed by a point particle, and that the expansion parameter is $\epsilon$, which counts powers of $m$ and $q$. On the other hand, in the particle view we have that the field of the charged particle is tidally deformed by the black hole, and the expansion parameter is formally $M$, to reflect the fact that $M/b^{\ell+1}$ measures the strength of the tidal field for a multipole of order $\ell$. In this section we compare the two constructions, show that they are indeed compatible, and give approximate expressions for the constants that were left undetermined in the preceding calculations. 

\subsubsection{Black-hole view} 

We begin with the metric and vector potential of the black-hole view. The metric is given in local $(t,s,\alpha,\phi)$ coordinates by Eq.~(\ref{metricBH_local}), in which $U = U_0 + \epsilon U_1 + \epsilon^2 U_2 + O(\epsilon^3)$, with $U_0$ given by Eq.~(\ref{01pot_local}), $U_1$ by Eq.~(\ref{U1}), and $U_2$ by Eq.~(\ref{U2sol}). The electrostatic potential is $\Phi = \epsilon \Phi_1 + \epsilon^2 \Phi_2 + O(\epsilon^3)$, with $\Phi_1$ given by Eq.~(\ref{01pot_local}), and $\Phi_2$ by Eq.~(\ref{Phi2sol}).

Noting that the expansion of $e^{-2U_0}$ in powers of $s$ begins with $(b-M)/(b+M)$, we find that inserting the potentials in the metric of Eq.~(\ref{metricBH_local}) produces an overall factor of $(b-M)/(b+M)$ in front of $dt^2$, and factors of $(b+M)/(b-M)$ in front of $ds^2$, $s^2\, d\alpha^2$, and $s^2 \sin^2\alpha\, d\phi^2$. These can be eliminated by a rescaling of the time and radial coordinates, 
\begin{equation} 
t = \sigma^{-1}\, \bar{t}, \qquad
s = \sigma\, \bar{s}, \qquad
\sigma := \sqrt{\frac{b-M}{b+M}}. 
\end{equation} 
With this notation we find that the metric of the black-hole view becomes 
\begin{equation}   
ds^2 = -\bar{V}\, d\bar{t}^2 + \bar{V}^{-1} e^{2\bar{\gamma}} (d\bar{s}^2 + \bar{s}^2\, d\alpha^2)
+ \bar{V}^{-1} \bar{s}^2\sin^2\alpha\, d\phi^2, 
\label{metric_match} 
\end{equation} 
where 
\begin{equation} 
\bar{V} := \sigma^{-2} e^{-2U}
= \sigma^{-2} e^{-2U_0} \biggl[ 1 - 2\epsilon U_1 + 2\epsilon^2 \bigl( U_1^2 - U_2 \bigr)
+ O(\epsilon^3)  \biggr]  
\end{equation} 
and $\bar{\gamma} = \gamma$. Because the electrostatic potential is (minus) the $t$-component of the vector potential, it is affected by a rescaling of the time coordinate. Defining $-\bar{\Phi} := A_{\bar{t}} = A_t (dt/d\bar{t})$, we have that 
\begin{equation} 
\bar{\Phi} = \sigma^{-1}\, \Phi
= \sigma^{-1} \Bigl[ \epsilon \Phi_1 + \epsilon^2 \Phi_2 + O(\epsilon^3) \Bigr]. 
\label{potential_match} 
\end{equation} 

We import our results from Sec.~\ref{sec:electric-charge}, and in anticipation of the comparison with expressions obtained from the particle view, we formally linearize them with respect to $M$. Setting $\epsilon = 1$, we have that  
\begin{subequations}
\begin{align} 
\bar{V} &= \frac{2m^2+q^2}{\bar{s}^2} 
+ \frac{1}{\bar{s}} \biggl[ -2m + \frac{4M}{b^2}(m^2+q^2) \cos\alpha \biggr]  
+ 1 - 2\mu_0 + \frac{M}{b^3} (2m^2 + q^2)   
\nonumber \\ & \quad \mbox{} 
- \frac{2M}{b^2} \biggl(2m + \frac{q^2}{b} \biggr) \cos\alpha 
- \frac{M}{b^3} (6m^2 + 5q^2) \cos^2\alpha + O(\bar{s}), \\  
\bar{\Phi} &= -\frac{mq}{\bar{s}^2} 
+ \frac{q}{\bar{s}} \biggl[ 1 - \frac{2mM}{b^2} \cos\alpha \biggr]
+ \frac{qM}{b^2} + \nu_0 \biggl( 1 + \frac{M}{b} \biggr) 
+ \frac{q M}{b^2} \biggl( 1 + \frac{m}{b} \biggr) \cos\alpha 
+ \frac{5mqM}{2 b^3} \cos^2\alpha + O(\bar{s}). 
\end{align} 
\end{subequations} 

\subsubsection{Particle view} 

The particle-view metric of Eq.~(\ref{metricPart}) is already of the form of Eq.~(\ref{metric_match}), with 
\begin{equation} 
\bar{V} = e^{-2\bar{U}} = f e^{-2\delta U} \simeq f (1 - 2\delta U). 
\end{equation} 
The electrostatic potential $\bar{\Phi} = \bar{\Phi}_0 + \delta \Phi$ refers to the same time coordinate as in Eq.~(\ref{potential_match}). The perturbations $\delta U$ and $\delta \Phi$ were obtained in Sec.~\ref{sec:PartView} as multipole expansions. To enable a comparison of these expressions to those of the black-hole view, we let $m \to \epsilon m$, $q \to \epsilon q$, expand through order $\epsilon^2$, and finally set $\epsilon = 1$. At this order, keeping $\bar{V}$ and $\bar{\Phi}$ accurate through order $\bar{s}^0$, we find that it is sufficient to keep terms up through $\ell = 2$ in the multipole expansion. We obtain
\begin{subequations} 
\begin{align} 
\bar{V} &= \frac{2m^2+q^2}{\bar{s}^2}
+ \frac{1}{\bar{s}} \biggl\{ -2m - 4 \bigl[ (m^2+q^2) u_1 + mq p_1 \bigr] \cos\alpha \biggr\} 
+ 1 + \biggl( \frac{8}{3} m^2 + q^2 \biggr) u_2 + \frac{7}{3} m q p_2 
\nonumber \\ & \quad \mbox{}
+ 2(2m u_1 + q p_1) \cos\alpha - \bigl[ (6m^2+5q^2) u_2 + 7mq p_2 \bigr] \cos^2\alpha
+ O(\bar{s}), \\ 
\bar{\Phi} &= -\frac{mq}{\bar{s}^2} + \frac{q}{\bar{s}} \biggl\{ 1 + ( 2m u_1 + q p_1 )\cos\alpha \biggr\} 
- \frac{5}{6} m q u_2 - \frac{1}{6}(3m^2 + q^2) p_2 - (q u_1 + m p_1)\cos\alpha 
\nonumber \\ & \quad \mbox{}
+ \biggl[ \frac{5}{2} m q u_2 + \frac{1}{2} (m^2 + 3q^2) p_2 \biggr] \cos^2\alpha
+ O(\bar{s}). 
\end{align} 
\end{subequations}

\subsubsection{Approximate determination of constants} 

A detailed comparison between the two versions of $\bar{V}$ and $\bar{\Phi}$ reveals that the expressions match provided that 
\begin{subequations} 
\begin{align} 
u_1 = -\frac{M}{b^2} + O(\epsilon^2), & \qquad p_1 = -\frac{qM}{b^3} + O(\epsilon^2), \\ 
u_2 = \frac{M}{b^3} + O(\epsilon^2), & \qquad p_2 = \frac{qM}{b^4} + O(\epsilon^2) 
\end{align} 
\end{subequations} 
and 
\begin{equation} 
\mu_0 = -\frac{m^2 M}{3b^3} + O(M^2), \qquad 
\nu_0 = -\frac{5mq M}{6b^3} + O(M^2). 
\end{equation} 
As expected, we find that $u_\ell \propto M/b^{\ell+1}$ measures the strength of the tidal field; $p_\ell$ is further suppressed by a factor of $q/b$. With these assignments, we find that
\begin{equation}
\bar{V}[\mbox{black-hole view}] = \bar{V}[\mbox{particle view}] + O(\epsilon^3, M^2) 
\end{equation}
and
\begin{equation}
\bar{\Phi}[\mbox{black-hole view}] = \bar{\Phi}[\mbox{particle view}] + \frac{qM}{b^2} + O(\epsilon^3, M^2).  
\end{equation}
The electrostatic potentials agree up to an irrelevant constant. This constant could have been included within $\nu_0$. We chose to exclude it, because being part of $\Phi_2$, $\nu_0$ must scale as $\epsilon^2$, while $qM/b^2$ is linear in $\epsilon$.   

We conclude that the black-hole and particle views do indeed return mutually compatible approximations for the gravitational and electrostatic potentials. 

\subsection{Absence of multipole moments} 

In Sec.~\ref{sec:second-order} we made the assertion that the gravitational and electrostatic potentials constructed in the black-hole view should not contain singular terms that would be associated with the presence of higher multipole moments. The assertion was justified on the grounds that the field equations do not feature distributional sources for these moments. Here we provide an additional argument in favor of the assertion.  

The physical basis for the assertion is the stipulation that the particle should be as spherical as possible. This requirement was incorporated in the particle view by taking the particle's unperturbed field to be described by the Reissner-Nordstr\"om solution to the Einstein-Maxwell equations. The particle's field, however, cannot be strictly spherical, because it is deformed by the presence of the black hole. While we have allowed for such a tidal deformation, we have properly ruled against intrinsic multipole moments by discarding the decaying solutions to the perturbation equations.  

The subtlety is whether the tidal deformation of the particle's field could induce terms in $U_2$ and $\Phi_2$ that {\it appear} to be associated with intrinsic moments, that is, terms that are singular in the limit $s \to 0$. It is easy to show that this cannot happen. 

To begin, consider $\delta U_1$ and $\delta \Phi_1$, the dipole contributions to the particle-view potentials, as listed in Eqs.~(\ref{tidal_multipoles}). The transformation from $r$ to $s$ is given by Eq.~(\ref{inverse_local}), and schematically we have that $r = s + \epsilon + \epsilon^2/s + O(\epsilon^3)$, where $\epsilon$ stands for either $m$ or $q$, and $\epsilon^2$ stands for either $m^2$, $mq$, or $q^2$. Incorporating this transformation, we find that through order $\epsilon^2$, $\delta U_1$ and $\delta \Phi_1$ cannot be more singular than $s^{-1}$, which is short of the $s^{-2}$ singularity that could be associated with an intrinsic dipole. Continuing this examination for $\delta U_2$ and $\delta \Phi_2$, the quadrupole contributions to the particle-view potentials, we find that these cannot be singular at all; the smallest power of $s$ that appears through order $\epsilon^2$ is $s^0$, and this is nowhere near the $s^{-3}$ singularity that would come with an apparent intrinsic quadrupole. Going to higher multipoles confirms the trend: the expressions for $\delta U_\ell$ and $\delta \Phi_\ell$, truncated through order $\epsilon^2$, have $s^{\ell - 2}$ as the smallest power of $s$, and this cannot match the $s^{-(\ell+1)}$ singularity of an apparent multipole moment. Notice that the argument is completely indifferent to the value of the amplitudes $u_\ell$ and $p_\ell$, so long as these do not scale with inverse powers of $\epsilon$. (The detailed matching exercise reveals that they do not.)

Our conclusion thus far is that a first-order tidal perturbation cannot produce terms in $U$ and $\Phi$ that could mimic those associated with intrinsic multipole moments. Another argument, based entirely on dimensional analysis, supports this conclusion, and allows us to extend it to nonlinear tidal perturbations.

The argument goes as follows. In geometrized units, a (mass or charge) multipole moment $d_\ell$ has a  dimension of length raised to the power $\ell+1$, where $\ell \geq 1$. Three independent length scales are available to make up this multipole moment: $\epsilon$, $M$, and $b$. If the multipole moment is to scale with $\epsilon^2$ and linearly with $M$ (for a first-order tidal deformation), we must have a relation of the form $d_\ell \sim \epsilon^2 M b^{\ell-2}$. This, however, does not match an expected scaling with $M$ and $b$ through the combination $M/b^{\ell+1}$, which would result if the apparent multipole moment were the result of a tidal deformation. The only way to reconcile these scalings would be to introduce another length scale, such as a particle radius, into the problem; in the absence of such a scale, we must conclude that the tidal deformation cannot induce the presence of apparent multipole moments.

The argument generalizes to higher order in an expansion in powers of the tidal interaction. To keep things specific, let us consider a second-order treatment of the tidal perturbation. In this situation the multipole moment continues to scale as $\epsilon^2$, but it is now proportional to $M^2$, and we have a relation of the form $d_\ell \sim \epsilon^2 M^2 b^{\ell-3}$. On the other hand, if this apparent multipole moment were the result of a second-order tidal interaction, we would expect a scaling with $M$ and $b$ through the combination $M^2/b^{\ell_1 + \ell_2 + 2}$, where $\ell_1$ and $\ell_2$ are the multipole orders of the first-order interaction. The composition of spherical harmonics implies that $\ell_1 + \ell_2 \geq \ell$, and it follows that the tidal scaling can never be a match for a required $\epsilon^2 M^2 b^{\ell-3}$. We conclude that a second-order tidal deformation cannot induce the presence of apparent multipole moments.

More thought along these lines reveals that the conclusion continues to apply at higher orders in the tidal interaction. The argument provides compelling evidence in favor of the assertion that the gravitational and electrostatic potentials cannot contain singular terms that would be associated with tidally-induced multipole moments.  

\bibliography{/Users/poisson/writing/papers/tex/bib/master}

\end{document}